\documentclass[sigconf,nonacm,authoryear]{acmart}
\usepackage{booktabs}
\usepackage{graphicx}

\usepackage{amsmath,amssymb,amsthm}
\usepackage{algorithm}
\usepackage[noend]{algpseudocode}
\usepackage[capitalise,noabbrev]{cleveref}
\usepackage{placeins}
\usepackage{tikz}
\usetikzlibrary{arrows.meta,positioning,fit,calc,shapes.geometric}
\graphicspath{{figs/}}

\newtheorem{proposition}{Proposition}
\newtheorem{lemma}{Lemma}

\newtheorem{theorem}{Theorem}
\newtheorem{example}{Example}
\theoremstyle{acmdefinition}
\newtheorem{definition}{Definition}
\theoremstyle{acmplain}
\title[Strategyproof Deep RL]{Strategyproof Deep Reinforcement Learning\\for Online Resource Allocation}

\author{Fariba Dehghan\textsuperscript{*},
  Sebastian Stein\textsuperscript{*},
  Vahid Yazdanpanah\textsuperscript{*},
  Stephanie Gauthier\textsuperscript{*}}
\affiliation{%
  \institution{\textsuperscript{*}University of Southampton}
  \country{UK}
    \{F.Dehghan, V.Yazdanpanah, S.Gauthier\}@soton.ac.uk,
    ss2@ecs.soton.ac.uk}

\begin{abstract}
Online resource allocation faces two coupled challenges: (1) allocating scarce resources efficiently under uncertain future demand, and (2) ensuring that agents report task valuations and requirements truthfully (strategyproofness).
Reinforcement Learning (RL) adapts allocation decisions through experience but does not necessarily ensure strategyproofness. Existing strategyproof RL uses a linear allocation rule, which limits the resource interactions it can represent. Neural mechanism design methods offer greater expressiveness, but provide only approximate incentive guarantees. 
Against this background, we introduce the first strategyproof deep RL mechanism for online resource allocation. It integrates three novel components: (i) a monotonic neural network that learns nonlinear interactions by construction (higher reported value cannot reduce acceptance, while a longer reported duration or larger resource requirement cannot increase it); (ii) a generalised critical payment: the lowest accepted value, with all other inputs fixed; and (iii) training by either RL or imitation of an existing rule. Together, monotonicity and the critical payment ensure strategyproofness and individual rationality, while the neural rule provides flexibility and imitation improves training efficiency. 
Empirically, on a four-resource stream with two partner pairs that a linear difference cannot represent, the neural rule achieves 97.8\% of an offline welfare upper bound, compared with 93.8\% for a linear learner and 89.0\% for first-come, first-served, closing 65\% and 80\% of the remaining optimality gap, respectively. Moreover, imitation reaches 97.5\% in only 261 seconds, about 89 times faster than RL.
\end{abstract}


\newcommand{\BibTeX}{\rm B\kern-.05em{\sc i\kern-.025em b}\kern-.08em\TeX}

\begin{document}

\maketitle 

%%%%%%%%%%%%%%%%%%%%%%%%%%%%%%%%%%%%%%%%%%%%%%%%%%%%%%%%%%%%%%%%%%%%%%%%
\section{Introduction}
\label{sec:intro}

In resource allocation problems, a limited capacity needs to be assigned among competing tasks. In the setting we study, each task has a value, a duration and a demand for one or more resource types. In the \emph{online} version we study, later arrivals are unknown, and the provider must accept or reject each task upon arrival. This structure appears when packing jobs onto shared compute~\cite{stein2020sprl,ghodsi2011dominant}, matching flexible electrical load to uncertain supply~\cite{stroehle2014online}, and serving electric vehicles at a constrained charging site~\cite{gerding2011ev,lee2021acn}. Accepting a task earns value but reserves capacity that a later, more valuable task may need. The first challenge is therefore to maximise social welfare, the total true value of completed tasks, without knowing future demand.
The second challenge is strategic reporting. Value, duration and demand are private, so a task owner may exaggerate value, understate the work it requires, or otherwise alter its report to obtain service or pay less. A mechanism is \emph{strategyproof} when reporting all three quantities truthfully maximises the task owner's utility, regardless of the other reports. For an owner who values only complete service, utility is its value minus its payment. A standard route to strategyproofness combines a monotone allocation rule with a critical payment: a higher value cannot hurt acceptance, a longer duration or greater demand cannot help it, and an accepted task pays the smallest value that the deployed rule accepts~\cite{myerson1981optimal,blumrosen2007combinatorial}. Our challenge is to retain this structure when the allocation rule is a nonlinear learned function whose threshold must be computed numerically.

Analytical online mechanisms provide strategyproof allocation and payment rules for electric vehicles, flexible loads, reusable resources and multi-resource allocation~\cite{gerding2011ev,stein2012precommit,stroehle2014online,gerding2016v2g,nejad2017online,zou2017game,dong2021nonmyopic,sikdar2021sequential,goko2022subsidy,aziz2020exchange,anandanarayanan2025truthful}. Their allocation rules, however, are specified in advance. Reinforcement learning (RL) and deep-RL methods, on the other hand, learn adaptive charging, pricing and resource-allocation policies~\cite{lee2021acn,wang2021rl,cao2022smart,zhao2022dynamic,hou2020rmd,gebrekidan2024drl,liu2025v2b}, but generally do not prove truthful reporting of value, duration and demand. Neural auction design often penalises empirical regret~\cite{duetting2019auctions}; other one-shot architectures enforce stronger guarantees for restricted auction domains~\cite{shen2019automated,duetting2024auctions,ma2026learning}, including exact incentive compatibility by representing the mechanism as a menu from which the buyer selects~\cite{shen2019automated}, and post-hoc tools verify fixed networks~\cite{curry2020certifying}. These methods address complementary parts of the problem, yet they do not directly provide a per-arrival critical payment for a learned sequential rule whose current acceptance changes future capacity. Without strategyproofness, even a high-welfare learned rule cannot be used as intended~\cite{stroehle2014online,stein2012precommit}. Owners can raise the reported value or change the duration or demand to obtain service that would otherwise be refused, so the mechanism decides on false information~\cite{stein2020sprl}; later arrivals are displaced, and the resulting schedule is inefficient because it is not based on true requests~\cite{nejad2017online,gerding2016v2g}.

The closest existing work is the strategyproof RL mechanism of~\citet{stein2020sprl}. It learns a linear difference between the predicted returns from accepting and rejecting, projects its coefficients after each update to preserve monotonicity, and obtains the critical value by solving a linear equation. That difference is linear in occupancy and task features, with a single occupancy coefficient shared across all tasks. As a result, one occupancy overlap of the following form can be represented that way, whereas two cannot. For example, when tasks require different combinations of resources, a shared occupancy weight can encode one overlap between a low-value class and a high-value class, but not two overlaps that need opposite occupancy weights (\cref{ex:partner}). 
By contrast, a later proximal policy optimisation (PPO) mechanism uses clustered past reports both to shape the training reward of a neural policy and to set a payment~\cite{mishra2023monotonic}. That shaping term acts only on reports seen in training, so the trained allocation need not be monotone in reported value, duration, and demand, and the payment is a cluster statistic rather than the smallest value the deployed rule accepts. 
We therefore need two rules: a learned allocation that represents these occupancy interactions, for welfare, and a critical payment that makes truthful reporting optimal. The second requires the first to remain monotone in every reported value, duration, and demand.

To this end, a smooth min--max (SMM) network~\cite{igel2024smm} constructs the accept--reject difference from nested smooth (log-sum-exp) minima and maxima of linear units, keeping the weights on reported coordinates positive. Nested min and max of increasing maps remain increasing, so the network can represent occupancy interactions while every permitted report keeps the required direction after every training update. We use this class because it is differentiable, hence trainable by RL or imitation, and monotone by construction rather than by a penalty on sampled reports~\cite{gupta2019incorporate}.

The resulting mechanism, Strategyproof SMM (SP-SMM), is this allocation rule together with its critical payment. Because the neural threshold generally has no closed-form expression, we enclose it by verified numerical evaluation and charge the resulting bound. The same construction also applies to tasks that may require several resource types, without a task-specific pricing formula. Finally, we train the same network in two ways. RL learns from social welfare feedback in a simulator. Imitation instead copies accept/reject decisions from a teacher that do not need to be strategyproof. Discarding the teacher before deployment is what makes that permissible. The deployed mechanism is the monotone network with its critical payment, so strategyproofness comes from the deployed network, not from the teacher. In addition, imitation can reduce training cost. We compare behaviour cloning (BC), dataset aggregation (DAgger), and disturbances for augmenting robot trajectories (DART)~\cite{ross2010imitation,ross2011dagger,laskey2017dart}.

\subsection*{Contributions}
\label{sec:contributions}

\begin{enumerate}
  \item SP-SMM learns general nonlinear occupancy interactions while remaining monotone in every reported quantity, and a verified critical payment makes the rule strategyproof and individually rational for tasks that may require several resource types.
  \item On the two-pair stream of \cref{ex:partner}, SP-SMM reduces the linear learner's optimality gap to the offline LP bound from 6.2\% to 2.2\%, closing 65\% of that gap, whereas linear rules remain competitive when a single occupancy weight suffices.
  \item The same mechanism can be trained by RL or imitation without changing its incentives; imitation reaches a 2.5\% optimality gap in 261 seconds, about 89 times faster than RL.
\end{enumerate}
The rest of this paper is as follows. \Cref{sec:model} specifies the setting, \Cref{sec:method} the mechanism, \Cref{sec:experiments} the experiments, and \Cref{sec:conclusion} concludes.

\section{System Model}
\label{sec:model}

This section specifies the allocation setting. We first model resource capacities, private types, reports, occupancy and the sequential Markov decision process (MDP) (\cref{sec:env}). We then define sequential allocation, agent utility, and strategyproofness and individual rationality (\cref{sec:incentives}). A table of the principal symbols is in \cref{app:notation}.

\subsection{Environment}
\label{sec:env}

We model a resource provider that offers $m$ resource types $r\in\{1,\ldots,m\}$ with capacities $\mathbf{b}=(b_1,\ldots,b_m)\in\mathbb{N}^{m}$, so resource $r$ has $b_r$ units. We represent time as discrete slots $t\in\{1,\ldots,T\}$ over a horizon of \(T\) slots, and we assume one task arrives per slot. We index tasks by arrival order $i$, and we represent the private type of task $i$ as
\begin{equation}
  \label{eq:type}
  \theta_i \;=\; (v_i,\ell_i,\mathbf{d}_i),
\end{equation}
with value $v_i\ge 0$, duration $\ell_i\in\mathbb{N}$, and per-slot demand $\mathbf{d}_i=(d_{i,1},\ldots,d_{i,m})\in\mathbb{N}_0^{m}$, where $\mathbb{N}_0=\{0,1,\ldots\}$ and $d_{i,r}$ is the requirement on resource $r$.

We assume a bounded type domain \(\Theta\) with \(v_{\max}>0\) and \(\ell_{\max}\),
\begin{equation}
  \label{eq:type-domain}
  \Theta
  \;=\;
  [0,v_{\max}]
  \times
  \{1,\ldots,\ell_{\max}\}
  \times
  \prod_{r=1}^{m}\{0,\ldots,b_r\}.
\end{equation} 

We further assume that only agent $i$ knows $\theta_i$, and the provider must accept or reject each task upon arrival, before later arrivals are known, and that the decision is irrevocable. The provider cannot measure value, and neither duration nor demand is verifiable before that decision, so we model the mechanism as acting on a \emph{report} \(\hat\theta_i=(\hat v_i,\hat\ell_i,\hat{\mathbf{d}}_i)\in\Theta\). We do not assume that \(\hat\theta_i = \theta_i\). By contrast, we do not consider strategic reporting of the arrival time. In particular, the provider observes the slot at which the task appears. This is realistic for time-sensitive tasks; for example, an electric vehicle already at a charging site~\cite{gerding2011ev,lee2021acn} or a compute job that must start when submitted~\cite{stein2020sprl}. When discussing the current arrival we drop the index \(i\) and write \((\hat v,\hat\ell,\hat{\mathbf{d}})\).

We represent the provider's state \(s\) by the current slot \(t\) and the remaining busy time \(o_{r,k}(s)\) of each unit \(k\in\{1,\ldots,b_r\}\) of each resource \(r\). We write \(f_r(s)\) for the number of those coordinates that are zero, so \(\mathbf f(s)=(f_1(s),\ldots,f_m(s))\in\mathbb{N}_0^m\) is the currently free capacity. We treat a report as \emph{feasible} when
\begin{equation}
  \label{eq:feasible}
  \hat d_{i,r}\le f_r(s)
  \quad\forall r\in\{1,\ldots,m\}
  \quad\text{and}\quad
  \hat\ell_i\le T-t+1.
\end{equation}

Consequently, every accepted task completes by slot \(T\), and infeasible reports are rejected. This \(s\) is therefore the state of an MDP~\cite{sutton2018reinforcement}: accepting a task changes occupancy for later slots, so each arrival is a sequential decision. The actions are accept and reject. Let \(I_r\subseteq\{k:o_{r,k}(s)=0\}\) with \(|I_r|=\hat d_{i,r}\) be the idle units of resource \(r\) assigned when the report is accepted, and \(I_r=\emptyset\) otherwise, \(\forall r\in\{1,\ldots,m\}\). The next slot then has time \(t+1\) and remaining busy times
\begin{equation}
\label{eq:occupancy-update}
o_{r,k}
\leftarrow
\begin{cases}
\max(\hat\ell_i-1,0) & \text{if }k\in I_r,\\
\max(o_{r,k}(s)-1,0) & \text{otherwise,}
\end{cases}
\end{equation}
\(\forall r\in\{1,\ldots,m\},\;\forall k\in\{1,\ldots,b_r\}\). This occupancy update is the MDP transition. The next report, however, is unknown.

\subsection{Mechanism, utility and incentives}
\label{sec:incentives}

Because types are private, the provider cannot allocate from \(\theta_i\) directly. The next definition is the sequential rule that turns a report into an allocation decision and a payment.
\begin{definition}[Mechanism]
\label{def:mech}
A mechanism is a map~\cite{myerson1981optimal,gerding2011ev,stein2020sprl}
\begin{equation}
  \label{eq:mech-maps}
  \mathcal{M}(\mathcal{E},\hat\theta_i,\hat\Theta_{<i})
  \;=\;
  \bigl(x(\hat\theta_i,\hat\Theta_{<i}),\,p(\hat\theta_i,\hat\Theta_{<i})\bigr),
\end{equation}
with \(x\in\{0,1\}\), \(p\ge 0\), and \(x=0\Longrightarrow p=0\).\footnote{We omit the arguments of \(x\) and \(p\) when they are clear from context.} The input is the environment \(\mathcal{E}=(m,\mathbf{b},T,\Theta,(o_{r,k}))\) of \cref{sec:env}, the current report \(\hat\theta_i\in\Theta\) of agent \(i\), and the history \(\hat\Theta_{<i}=(\hat\theta_1,\ldots,\hat\theta_{i-1})\in\Theta^{i-1}\). The output is acceptance (\(x=1\)) or rejection (\(x=0\)) and agent \(i\)'s payment. We take \(\mathcal{M}\) to be deterministic, so occupancy \(o_{r,k}\) at arrival \(i\) is determined by \(\hat\Theta_{<i}\) (\cref{eq:occupancy-update}). With \(\mathcal{E}\) as in \cref{sec:env}, \(x\) and \(p\) therefore depend only on \(\hat\theta_i\) and \(\hat\Theta_{<i}\).
\end{definition}

We denote a report as \emph{covering} the true type when its reported duration and demands are at least the corresponding true quantities,
\begin{equation}
  \label{eq:cover}
  c(\hat\theta_i,\theta_i)
  \;=\;
  \mathbf{1}\bigl[\hat\ell_i\ge\ell_i,\;
    \hat d_{i,r}\ge d_{i,r}\ \forall r\in\{1,\ldots,m\}\bigr],
\end{equation}
where $\mathbf{1}[\cdot]$ denotes the indicator of the event in brackets. An understated duration or demand (where $c(\hat\theta_i,\theta_i)=0$) then fails to complete the true task~\cite{stein2020sprl}. We take social welfare to be the total true value of accepted covering reports, and we maximise $W$ rather than revenue because the provider allocates a shared resource~\cite{stein2020sprl}:
\begin{equation}
  \label{eq:welfare}
  W
  \;=\;
  \sum_{i}
  x(\hat\theta_i,\hat\Theta_{<i})\,
  c(\hat\theta_i,\theta_i)\,
  v_i.
\end{equation}

We assume that agent $i$ maximises utility when choosing a report~\cite{myerson1981optimal}, and with the covering convention we represent that payoff as:
\begin{definition}[Agent utility]
\label{def:utility}
The utility of agent $i$ is
\begin{equation}
  \label{eq:utility}
  U_i(\hat\theta_i,\theta_i,\hat\Theta_{<i})
  \;=\;
  x(\hat\theta_i,\hat\Theta_{<i})\,
  c(\hat\theta_i,\theta_i)\,
  v_i
  \;-\;
  p(\hat\theta_i,\hat\Theta_{<i}).
\end{equation}
\end{definition}
Under this utility, an accepted report that understates duration or demand yields no value but still incurs a payment, whereas a rejection yields \(U_i=0\), since \(x=0\) and \(x=0\Longrightarrow p=0\). A misreport is profitable exactly when it raises $U_i$.

A mechanism is strategyproof when truthful reporting is a dominant strategy, and individually rational when a truthful agent is never worse off than not participating~\cite{myerson1981optimal}:
\begin{definition}[Strategyproofness and individual rationality]
\label{def:sp}
$\mathcal{M}$ is strategyproof if $U_i(\theta_i,\theta_i,\hat\Theta_{<i})
  \;\ge\;
  U_i(\hat\theta_i,\theta_i,\hat\Theta_{<i})$, $\forall i,\;
  \forall\theta_i,\;
  \forall\hat\theta_i,\;
  \forall\hat\Theta_{<i}.$
It is individually rational if $U_i(\theta_i,\theta_i,\hat\Theta_{<i})
  \;\ge\;
  0$, $\forall i,\;
  \forall\theta_i,\;
  \forall\hat\Theta_{<i}.$
\end{definition}

Because we assume irrevocable decisions before later arrivals (\cref{sec:env}), future reports cannot affect the current decision or payment, so these definitions need only the preceding history. Strategyproofness matters because \(x\) is decided on reports, while \(W\) uses true values \(v_i\). The two align only under truthful reporting. This can be achieved using the characterisation of~\citet{blumrosen2007combinatorial}. Specifically, with our arrival model and the covering convention, agents are single-minded in their sense, so a monotone allocation rule together with the critical-value payment (i.e., the lowest reported value the rule accepts) is strategyproof and individually rational~\cite{myerson1981optimal,blumrosen2007combinatorial}.
\Cref{sec:method} constructs that pair as SP-SMM.

%% ===========================================================================
\section{Strategyproof Smooth Min--Max}
\label{sec:method}

This section constructs SP-SMM from action values on the MDP of \cref{sec:env} (\cref{sec:mdp}), a monotone network (\cref{sec:smm}), and a critical payment (\cref{sec:invert}). 

\subsection{Action values}
\label{sec:mdp}

The realised welfare that follows acceptance or rejection cannot be evaluated exactly (\cref{sec:env}). We therefore learn estimates of the expected return (in terms of welfare) for both actions from sampled episodes~\cite{sutton2018reinforcement}, giving us a direct value for each action, rather than outputting only an accept/reject probability. During training, exploration provides experience of both actions and the occupancy, i.e., state, they produce. Q-learning would retain the action actually taken, its reward and the resulting occupancy, but would bootstrap from the highest estimated continuation at the next state. On-policy SARSA instead bootstraps from the next action selected by the behaviour policy. We use SARSA to evaluate continuation under that policy, including its exploratory choices. To keep those bootstrap targets fixed within each fitting iteration, we collect transitions, compute targets from a frozen network snapshot, and fit the network to those targets in batches; we then refresh the targets. That adapts the fitted-batch scheme of fitted Q-iteration~\cite{ernst2005tree,riedmiller2005nfq} to SARSA (\cref{alg:sarsa}).

The components of this learner, and their roles, are as follows. First, the state $s$ is the occupancy record of \cref{sec:env} together with the current report, so the action-value can depend on both free capacity and the request. Second, the action $a$ is the allocation \(x\): \(a=\mathrm{acc}\) if \(x=1\) and \(a=\mathrm{rej}\) if \(x=0\). The transition remains the occupancy update of \cref{eq:occupancy-update}. Consequently, the reward's role is to credit remaining welfare at the accept/reject decision. The action-value \(Q_w(s,a)\) then estimates the undiscounted return from that decision.
We record these objects as follows.
\begin{definition}[Action values and features]
\label{def:q-features}
Let \(w\) denote the network parameters, and let \(v_{\max}\) and \(\ell_{\max}\) be the bounds in \(\Theta\). The immediate reward and the undiscounted return from slot \(t\) are
\begin{equation}
  \label{eq:reward}
  R(s,a)
  \;=\;
  \mathbf{1}[a=\mathrm{acc}]\,\hat v,
  \qquad
  J_t
  \;=\;
  \sum_{k=t}^{T} R(s_k,a_k).
\end{equation}
Under truth-telling, \(\mathbb{E}[J_t]\) is the expected social welfare from slot \(t\) onward, and \(Q_w(s,a)\) is its estimate given \(s\) and \(a\). The accept--reject action-value difference and the greedy policy on feasible reports are
\begin{equation}
  \label{eq:action-value-difference}
  \Delta_w(s)=Q_w(s,\mathrm{acc})-Q_w(s,\mathrm{rej}),
  \quad
  \pi_w(s)=\mathrm{acc}\iff\Delta_w(s)\ge0.
\end{equation}
When \(\Delta_w\) equals the true difference of expected returns, \(\pi_w\) maximises expected social welfare~\cite{sutton2018reinforcement}. With remaining busy times ordered \(o_{r,1}(s)\le\cdots\le o_{r,b_r}(s)\), the normalised occupancy and the feature map are
\begin{equation}
  \label{eq:occupancy-feat}
  o(s)
  \;=\;
  \bigl(o_{r,k}(s)/\max(\ell_{\max}-1,1)\bigr)_{r,k},
\end{equation}
\begin{equation}
  \label{eq:features}
  \phi(s)
  \;=\;
  \bigl(
    1,\;
    t/T,\;
    \hat v/v_{\max},\;
    \hat\ell/\ell_{\max},\;
    \hat d_1/b_1,\;\ldots,\;\hat d_m/b_m,\;
    o(s)
  \bigr).
\end{equation}
Let \(\phi_j\) be the \(j\)-th coordinate of \(\phi(s)\); each \(\phi_j\in[0,1]\). This normalisation places elements on a common scale without changing their order or introducing interactions between them. In particular, we write \(u=\hat v/v_{\max}\in[0,1]\) and write \(\phi(s,u)\) for \(\phi(s)\) with the normalised-value coordinate set to \(u\). In addition, context \(z(s)=(1,t/T,o(s))\) collects time and occupancy.
\end{definition}
Because the agent chooses only the report \((\hat v,\hat\ell,\hat{\mathbf{d}})\), only those coordinates are constrained:
\begin{definition}[Monotone $\Delta_w$]
\label{def:monoA}
\(\Delta_w\) is monotone if, with time and occupancy held fixed,
\begin{equation}
  \label{eq:monoA}
  \hat v'\ge\hat v,\quad
  \hat\ell'\le\hat\ell,\quad
  \hat d'_r\le\hat d_r\ \forall r
  \quad\Longrightarrow\quad
  \Delta_w(\hat v',\hat\ell',\hat{\mathbf d}')
  \ge
  \Delta_w(\hat v,\hat\ell,\hat{\mathbf d})
\end{equation}
for every permitted pair of reports and every parameter vector \(w\).
\end{definition}
By contrast, time and occupancy enter only through \(z(s)\), which \cref{def:monoA} holds fixed.

We therefore need a representation of \(\Delta_w\) that satisfies \cref{def:monoA} while still letting occupancy change the accept/reject comparison. Because a shared baseline on \(s\) cancels in \(\Delta_w\), we parameterise \(\Delta_w\) directly as a difference network on the constrained report and unrestricted \(z(s)\). Occupancy may then interact with the report, rather than enter as one shared linear weight. \Cref{sec:smm} constructs that network. A linear \(\Delta_w\) with occupancy weights shared across tasks is a special case of the same comparison; \cref{ex:partner} is a setting that special case cannot represent.
\begin{example}[Two partner pairs]
\label{ex:partner}
Low-value tasks C need only \(r_1\) and valuable tasks A need \(r_1\) and \(r_2\) together: we call this a partner pair, because accepting C can block A only while the partner resource \(r_2\) remains free. A second partner pair is low-value D on \(r_4\) with valuable B on \(r_3\) and \(r_4\). Later tasks are unknown. Keep \(r_1\) and \(r_4\) free. The welfare-maximising decisions then reverse with occupancy:
\begin{center}
\small
\begin{tabular}{@{}lcc@{}}
\toprule
Occupancy & C & D \\
\midrule
\(r_2\) busy, \(r_3\) free & accept & reject \\
\(r_2\) free, \(r_3\) busy & reject & accept \\
\bottomrule
\end{tabular}
\end{center}
Thus the same report C is accepted when \(r_2\) is busy and often rejected when \(r_2\) is free, so only \(z(s)\) differs. Although task-specific intercepts can shift C relative to D, a linear \(\Delta_w\) with occupancy weights shared across tasks cannot represent both rows: the first places a greater weight on \(r_2\) than on \(r_3\), whereas the second places the opposite weight. A product of C's demand with occupancy of \(r_2\) would restore the four decisions; however, it is not a coordinate of that linear difference.
\end{example}
\Cref{alg:sarsa} fits \(w\) from these objects.
\begin{algorithm}[t]
  \caption{Fitted SARSA on the allocation MDP}
  \label{alg:sarsa}
\small
\begin{algorithmic}[1]
  \Require $\varepsilon\in[0,1]$, $w$, $N,B\in\mathbb{N}$
  \Function{Select}{$s,\varepsilon,w$}
    \If{$\hat d_r>f_r(s)$ for some $r$, or $\hat\ell>T-t+1$}
      \State \Return $(\mathrm{rej},\,0)$
      \label{ln:sarsa-infeas}
    \EndIf
    \State sample $\xi\sim\mathrm{Unif}[0,1]$
    \If{$\xi<\varepsilon$}
      \State sample $a\sim\mathrm{Unif}\{\mathrm{acc},\mathrm{rej}\}$
      \State \Return $(a,\,1)$
      \label{ln:sarsa-explore}
    \Else
      \State \Return $(\pi_w(s),\,0)$
      \label{ln:sarsa-greedy}
    \EndIf
  \EndFunction
  \State $\mathcal{D}\leftarrow\emptyset$
  \For{$n=1,\ldots,N$}
    \State $(o_{r,k})\leftarrow 0\quad\forall r,k$
    \label{ln:sarsa-init}
    \For{$t=1,\ldots,T$}
      \State observe $\hat\theta=(\hat v,\hat\ell,\hat{\mathbf{d}})$
      \State $s\leftarrow\bigl(t,(o_{r,k}),\hat\theta\bigr)$
      \label{ln:sarsa-state}
      \State $\phi(s)\leftarrow$ \cref{eq:features},\; $z(s)\leftarrow(1,t/T,o(s))$
      \label{ln:sarsa-phi}
      \State $(a,e)\leftarrow\textsc{Select}(s,\varepsilon,w)$
      \label{ln:sarsa-act}
      \State $R\leftarrow (1-e)\,\mathbf{1}[a=\mathrm{acc}]\,\hat v$
      \label{ln:sarsa-reward}
      \If{$t<T$}
        \State $(o_{r,k})\leftarrow$ \cref{eq:occupancy-update}
        \label{ln:sarsa-trans}
        \State observe $\hat\theta'$; $s'\leftarrow\bigl(t+1,(o_{r,k}),\hat\theta'\bigr)$
        \State $(a',e')\leftarrow\textsc{Select}(s',\varepsilon,w)$
        \label{ln:sarsa-next}
      \Else
        \State $s'\leftarrow\bot$; $a'\leftarrow\bot$
        \label{ln:sarsa-term}
      \EndIf
      \State $\mathcal{D}\leftarrow\mathcal{D}\cup\{(s,a,R,s',a')\}$
      \label{ln:sarsa-store}
    \EndFor
    \If{$n\bmod B=0$}
      \ForAll{$(s,a,R,s',a')\in\mathcal{D}$}
        \State $y\leftarrow R$ if $s'=\bot$, else $y\leftarrow R+Q_w(s',a')$
        \label{ln:sarsa-target}
      \EndFor
      \State hold $\{y\}$ fixed and update $w$
      \label{ln:sarsa-update}
    \EndIf
  \EndFor
  \State $\varepsilon\leftarrow 0$
  \label{ln:sarsa-deploy}
\end{algorithmic}
\end{algorithm}
Line~\ref{ln:sarsa-phi} of \cref{alg:sarsa} supplies \(\phi(s)\) and \(z(s)\). Line~\ref{ln:sarsa-target} is
\begin{equation}
  \label{eq:sarsa}
  R(s,a)+Q_w(s',a'),
\end{equation}
when \(s'\neq\bot\), with \(a'\) the stored next action (line~\ref{ln:sarsa-store}); otherwise, line~\ref{ln:sarsa-term} sets \(y=R\) at a terminal slot. In addition, line~\ref{ln:sarsa-reward} stores \(R=0\) when \(e=1\), so the randomised training policy is not the deployed mechanism. For this reason, line~\ref{ln:sarsa-deploy} sets \(\varepsilon=0\).

\subsection{Smooth min--max network}
\label{sec:smm}

We represent \(\Delta_w\) with the SMM network of~\citet{igel2024smm}. Because \cref{def:monoA} requires \(\Delta_w\) to decrease in duration and demand, we first map those coordinates by $\phi_j\mapsto 1-\phi_j$, which is valid because \cref{eq:features} places them in $[0,1]$. After that map, every report coordinate is increasing, so the units below need only positive weights. The core consists of $K$ groups of $H$ linear units, indexed by $k\in\{1,\ldots,K\}$ and $h\in\{1,\ldots,H\}$. Holding coordinates other than $u$ fixed, we take unit $(k,h)$ to be
\begin{equation}
  \label{eq:smm-unit}
  \zeta_{kh}(u)
  \;=\;
  a_{kh}+w_{kh}u,
  \qquad
  w_{kh}>0,
\end{equation}
with intercept $a_{kh}\in\mathbb{R}$. In addition, we store an unconstrained number for each weight attached to a report coordinate and exponentiate it, so the weight remains positive after every optimiser update. The unnormalised log-sum-exp maps are
\begin{equation}
  \label{eq:softextrema}
  \operatorname{smax}(\boldsymbol{\zeta})
  =\log\sum_h \exp(\zeta_h),
  \qquad
  \operatorname{smin}(\boldsymbol{\zeta})=-\operatorname{smax}(-\boldsymbol{\zeta}).
\end{equation}
We then take a smooth maximum over the units of each group and a smooth minimum over groups. With learned scale $\beta\in\mathbb{R}$ and offset $\gamma\in\mathbb{R}$,
\begin{equation}
  \label{eq:smm-core}
  f_{\mathrm{SMM}}(u)
  \;=\;
  \mathrm{e}^{-\beta}\,
  \operatorname{smin}_{k\le K}\operatorname{smax}_{h\le H}\zeta_{kh}(u)
  \;+\;
  \gamma.
\end{equation}
We also represent context by an unconstrained multilayer perceptron (MLP) with logistic activation that reads only $z(s)$ and returns a value $f_{\mathrm{ctx}}$~\cite{gupta2019incorporate}. We write $f_w(u)=f_{\mathrm{SMM}}(u)+f_{\mathrm{ctx}}$.
Finally, we apply a bounded output head to \(f_w\). We calibrate a centre $\mu$ and scale $\nu>0$ once at initialisation and then keep them fixed. For \(F_s(u)=\Delta_w(\phi(s,u))\), a learned positive gain \(\lambda\le\lambda_{\max}\) and offset \(b_0\) give
\begin{equation}
  \label{eq:bounded-difference}
  F_s(u)
  =
  \operatorname{clip}_{[-2\lambda_{\max},\,2\lambda_{\max}]}
  \left[
    \lambda\tanh\bigl((f_w(u)-\mu)/\nu\bigr)+b_0
  \right].
\end{equation}
\begin{lemma}[SMM monotonicity]
\label{lem:smm-mono}
Every parameter vector \(w\) of \cref{eq:smm-core,eq:bounded-difference} yields a monotone \(\Delta_w\) in the sense of \cref{def:monoA}.
\end{lemma}
\begin{proof}
After the duration--demand flip, every report coordinate is increasing and every report weight is positive, so each unit \(\zeta_{kh}\) increases in the report. Nested \(\operatorname{smin}\) and \(\operatorname{smax}\) of increasing maps remain increasing~\cite{igel2024smm}. The context term \(f_{\mathrm{ctx}}\) does not depend on the report, and \(\tanh\), scaling by \(\lambda>0\), and clipping are non-decreasing. The composition therefore satisfies \cref{def:monoA} for every \(w\).
\end{proof}
In particular, training may change the shape of the rule but cannot reverse a required report direction.
\begin{proposition}[Coefficient projection]
\label{prop:proj}
If \(\Delta_w\) is linear in \(\phi\), then \(\Delta_w\) satisfies \cref{def:monoA} if and only if the coefficients of normalised value, duration and demand have the required signs. The same coordinate-wise sign repair does not imply \cref{def:monoA} for a general MLP.
\end{proposition}
\begin{proof}
For a linear map the partial derivative in each coordinate of \(\phi\) is the corresponding coefficient, so \cref{def:monoA} holds at every report if and only if the value coefficient is nonnegative and the duration and demand coefficients are nonpositive. Projecting those coefficients onto the corresponding half-lines after each update restores the signs~\cite{stein2020sprl}. A general MLP has input-dependent derivatives, so clipping a first-layer weight does not control \(\partial\Delta_w/\partial\phi_j\) at every report.
\end{proof}
A monotonicity or regret penalty, on the other hand, only discourages violations at sampled reports; an unsampled report may still violate the required order~\cite{gupta2019incorporate}.

\subsection{Computing the critical payment}
\label{sec:invert}

The same smoothness that yields \cref{lem:smm-mono} also means \(F_s\) has no algebraic inverse: smooth extrema mix every unit and the \(\tanh\) head is nonlinear, whereas a hard min--max network of the same groups would have a closed form~\cite{sill1997monotonic}. We therefore invert the trained map itself: the complete action-value difference in \cref{eq:bounded-difference}. In particular, with duration and demand held fixed, that map is non-decreasing in \(u\in[0,1]\). The mathematical threshold uses a candidate \(u\) in place of the current reported value,
$u^\star(s)
  \;=\;
  \inf\{u\in[0,1]:F_s(u)\ge0\}$, $p^\star(s)=v_{\max}u^\star(s)$.
Thus, if \(F_s(0)\ge0\), then \(p^\star=0\). If \(F_s(1)<0\), no permitted value is accepted and we write \(p^\star=\infty\). Otherwise, the threshold lies between a rejecting value and an accepting value. Importantly, the search uses the candidate \(u\), not the current reported value. \Cref{alg:price-first} implements three stages. First, we evaluate candidate values with verified arithmetic. We then search until we have a proved bracket around \(p^\star\). Finally, we correct the upper endpoint and use that value as the payment.

\begin{algorithm}[t]
  \caption{Verified critical value before allocation}
  \label{alg:price-first}
\small
\begin{algorithmic}[1]
    \Procedure{DecideAndPrice}{$s,(\hat v,\hat\ell,\hat{\mathbf d}),F_s$}
      \If{\(\hat{\mathbf d}\nleq\mathbf f(s)\) or \(\hat\ell>T-t+1\)} \Return \((x,p)=(0,0)\)
      \label{ln:infeas}
      \EndIf
      \State \(p_{\mathrm{lo}}\leftarrow p_{\mathrm{hi}}\leftarrow\bot\)
      \For{\(P\in\{64,128,256\}\) bits}
        \State enclose \(F_s(0),F_s(1)\) at \(P\) bits
        \label{ln:eval}
        \If{\(F_s(1)<0\) verified} \Return \((x,p)=(0,0)\)
        \label{ln:none}
        \EndIf
        \If{\(F_s(0)\ge0\) verified}
        \label{ln:all}
          \State \(p_{\mathrm{lo}}\leftarrow p_{\mathrm{hi}}\leftarrow0\);
            \textbf{break}
        \label{ln:all-set}
        \EndIf
        \If{\(F_s(0)<0\le F_s(1)\) unverified} \textbf{continue}
          \label{ln:bump}
        \EndIf
        \State \(u_{\mathrm{lo}}\leftarrow0,\ u_{\mathrm{hi}}\leftarrow1\)
        \label{ln:init}
        \While{\(v_{\max}(u_{\mathrm{hi}}-u_{\mathrm{lo}})>\tau\)}
        \label{ln:while}
          \State \algorithmicif\ SMM \algorithmicthen\ \(u\leftarrow\) Newton \algorithmicelse\ \(u\leftarrow\) midpoint
          \label{ln:step}
          \If{\(F_s(u)<0\)} \(u_{\mathrm{lo}}\leftarrow u\)
          \label{ln:update}
          \ElsIf{\(F_s(u)\ge0\)} \(u_{\mathrm{hi}}\leftarrow u\)
          \label{ln:updatehi}
          \Else\ \textbf{continue} at next \(P\)
          \label{ln:restart}
          \EndIf
        \EndWhile
        \State \(p_{\mathrm{lo}}\leftarrow v_{\max}u_{\mathrm{lo}}\), \(p_{\mathrm{hi}}\leftarrow v_{\max}u_{\mathrm{hi}}\); \textbf{break}
        \label{ln:record}
      \EndFor
      \If{\(p_{\mathrm{hi}}=\bot\)} \Return unresolved error
      \label{ln:fail}
      \EndIf
      \State \(q\leftarrow\min\{v_{\max},\lceil p_{\mathrm{hi}}+\tau\operatorname{rank}(\hat\ell,\hat{\mathbf d})\rceil_{\mathbb F}\}\)
      \label{ln:q}
      \State \(x\leftarrow\mathbf{1}[\hat v\ge q]\), \(p\leftarrow x q\); \Return \((x,p)\)
      \label{ln:acc}
      \label{ln:rej}
    \EndProcedure
\end{algorithmic}
\end{algorithm}
We start by evaluating the network with Arb ball arithmetic~\cite{johansson2017arb}. The weights in a checkpoint are finite binary numbers, but ordinary floating-point evaluation need not return the exact real value of \(F_s(u)\). Rounding matters most near zero, where the accept/reject decision changes. For this reason, we represent a ball by the interval $B=[m_B-\rho_B,m_B+\rho_B]$, $\rho_B\ge0$, where \(m_B\) is its midpoint and \(\rho_B\) its radius. Outward rounding after every network operation guarantees that the exact real-valued result for the stored checkpoint lies in this interval~\cite{johansson2017arb}. The guarantee is deterministic; the ball is not a confidence interval. In particular, a negative upper endpoint proves rejection, and a nonnegative lower endpoint proves acceptance. If the lower endpoint is negative and the upper endpoint is nonnegative, the sign remains unresolved. Line~\ref{ln:eval} of \cref{alg:price-first} evaluates these balls at \(u=0\) and \(u=1\). Thus, if \(F_s(1)<0\) is verified, line~\ref{ln:none} rejects every permitted value; if \(F_s(0)\ge0\) is verified, lines~\ref{ln:all}--\ref{ln:all-set} set the threshold to zero. Arb verifies these evaluations but does not search for the threshold.

We then search for that threshold. After classifying the two endpoints, we either handle an endpoint case or start from the full interval \(I=[0,1]\) with the invariant
\begin{equation}
  \label{eq:bracket-invariant}
  F_s(u_{\mathrm{lo}})<0,
  \qquad
  F_s(u_{\mathrm{hi}})\ge0.
\end{equation}
Because \(F_s\) is non-decreasing, \(u^\star\in[u_{\mathrm{lo}},u_{\mathrm{hi}}]\). Line~\ref{ln:init} of \cref{alg:price-first} initialises this bracket. At each step we evaluate a new candidate with Arb. By default we take the midpoint: a verified negative sign moves the lower endpoint, and a verified nonnegative sign moves the upper endpoint (line~\ref{ln:update}). Each successful bisection halves the interval, so after \(n\) steps its normalised width is at most \(2^{-n}\). If a sign is unresolved, line~\ref{ln:restart} restarts the same search at \(128\) and then \(256\) bits (line~\ref{ln:bump} does the same when an endpoint sign cannot be verified). Higher precision nevertheless reduces arithmetic uncertainty but does not move either bracket endpoint by itself. If the endpoint signs or requested width still cannot be verified, line~\ref{ln:fail} returns an unresolved error rather than substituting an ordinary floating-point result.

In addition, for SMM we may try interval Newton (line~\ref{ln:step}), because the SMM core is smooth along the value axis~\cite{igel2024smm}. On a bracket where the output clamp is inactive, Arb can enclose the derivative of the complete action-value difference throughout the bracket. When that interval is strictly positive, interval Newton proposes a smaller root interval~\cite{nocedal2006numerical}. We use that proposal only if it is at least twice as narrow and fresh endpoint evaluations restore the rejecting--accepting invariant; otherwise we bisect. Newton therefore speeds up SMM but never replaces the bracketing proof. We also write the interval form in \cref{app:cert-arithmetic}.
We stop the search when the corresponding critical value interval has width at most
\(\tau=10^{-6}\) (line~\ref{ln:while}):
\begin{equation}
  \label{eq:root-enclosure}
  p_{\mathrm{lo}}
  =v_{\max}u_{\mathrm{lo}}
  \;\le\;
  p^\star
  \;\le\;
  v_{\max}u_{\mathrm{hi}}
  =p_{\mathrm{hi}},
  \qquad
  p_{\mathrm{hi}}-p_{\mathrm{lo}}\le\tau.
\end{equation}
Line~\ref{ln:record} records these endpoints. We convert each endpoint back to binary64 with outward rounding and check the result. Thus \(p_{\mathrm{hi}}\) is a proved upper bound on the mathematical network threshold, not an exact symbolic root. By contrast, a midpoint estimate would not give this one-sided guarantee.

Finally, we turn \(p_{\mathrm{hi}}\) into the deployed payment. However, two independently rounded upper bounds can appear in the wrong order when their exact thresholds differ by less than \(\tau\). For the integer duration and demands in \cref{eq:type}, we define the report rank $\operatorname{rank}(\hat\ell,\hat{\mathbf d})
  =(\hat\ell-1)+\sum_{j=1}^{m}\hat d_j.$
The rank is an ordering device, not a measure of resource cost or welfare. It counts unit increases from the least-demanding report \((\hat\ell,\hat{\mathbf d})=(1,\mathbf 0)\). 
Let \(\lceil z\rceil_{\mathbb F}\) denote the smallest binary64 number no smaller than \(z\). We take the deployed critical value to be
\begin{equation}
  \label{eq:deployed-critical}
  q(s)
  =
  \min\!\left\{
    v_{\max},\;
    \left\lceil
    p_{\mathrm{hi}}(s)+\tau\operatorname{rank}(\hat\ell,\hat{\mathbf d})
    \right\rceil_{\mathbb F}
  \right\}.
\end{equation}
Line~\ref{ln:q} of \cref{alg:price-first} computes this \(q\). The formula uses \(p_{\mathrm{hi}}\), duration and demand, but not \(\hat v\). We add the stored binary inputs exactly and round upward once; let \(\rho_q\) be that final increase. Because \(p_{\mathrm{hi}}\le p^\star+\tau\), $  0\le q-p^\star
  \le
  \bigl(\operatorname{rank}(\hat\ell,\hat{\mathbf d})+1\bigr)\tau+\rho_q$,
when the cap does not bind; the cap can only reduce the difference while remaining at least \(p^\star\). Here \(0\le\rho_q<\operatorname{ulp}(v_{\max})\). On the partner benchmark, \(\operatorname{rank}\le19\), so the two terms are at most \(2\cdot10^{-5}\) and \(4\cdot10^{-15}\) value units.
We then accept and charge using this \(q\). Line~\ref{ln:infeas} first rejects infeasible reports. Lines~\ref{ln:acc}--\ref{ln:rej} compare \(q(s)\) with the original value report, not a value reconstructed after normalisation, and charge \(q(s)\) upon acceptance. If we verify that no permitted value is accepted, we define \(q=+\infty\). On a feasible report (\cref{eq:feasible}) the deployed allocation is therefore
\begin{equation}
  \label{eq:cv}
  x(s)
  =
  \mathbf{1}\bigl[\hat v\ge q(s)\bigr],
\end{equation}
and \(x(s)=0\) otherwise.
An accepted task pays \(p(s)=q(s)\). In short, the network threshold \(p^\star\), the verified interval \([p_{\mathrm{lo}},p_{\mathrm{hi}}]\), and the deployed value \(q\) are different objects: \(p_{\mathrm{hi}}\) is a proved upper bound on \(p^\star\), and \(q\) is the number the implemented rule actually uses.
Altogether, the incentive properties of this rule are \cref{thm:sp}.

\begin{theorem}[Strategyproofness and individual rationality]
\label{thm:sp}
Assume integer duration and demand reports and a monotone \(\Delta_w\) (\cref{def:monoA}). This hypothesis holds for every SMM parameter vector (\cref{lem:smm-mono}) and, for a linear model, exactly after coefficient projection (\cref{prop:proj}). Suppose that, for every feasible report, the calculation verifies either \cref{eq:root-enclosure} or that no permitted value is accepted. Then the mechanism defined by \eqref{eq:deployed-critical} and \eqref{eq:cv}, taking \(q=+\infty\) when no permitted value is accepted, is strategyproof and individually rational for every \(w\).
\end{theorem}

\noindent\emph{Proof.}
For fixed duration and demand, \(q\) is computed after replacing the reported value by the search variable \(u\); hence, it is independent of \(\hat v\). Equation~\eqref{eq:cv} accepts exactly when \(\hat v\ge q\) and charges \(q\), so \(q\) is the critical value of the deployed rule. Raising the reported value cannot hurt acceptance.
Now compare two duration--demand reports. Report \(2\) has weakly greater duration and demand in every coordinate than report \(1\), with at least one strict inequality. First suppose both have finite thresholds. Let \(G_i=p_{\mathrm{hi},i}+\tau\operatorname{rank}_i\) denote the exact dyadic sum before its final conversion to binary64. Report monotonicity gives \(p^\star_2\ge p^\star_1\), while \(\operatorname{rank}_2\ge\operatorname{rank}_1+1\). The enclosure guarantee then gives
$G_2\ge p^\star_2+\tau\operatorname{rank}_2\ge p^\star_1+\tau(\operatorname{rank}_1+1) \ge G_1.$
Rounding upward to the least binary64 number is a non-decreasing map, and taking the minimum with the same \(v_{\max}\) is also non-decreasing. Therefore \(q_2\ge q_1\). If report 2 has no crossing, then \(q_2=+\infty\). If report 1 has no crossing, report monotonicity forces report 2 to have no crossing as well. These cases preserve the same order. A shorter duration or smaller demand also cannot destroy feasibility. The deployed allocation is therefore monotone in value, duration, and every demand coordinate.

For example, take two resources and the exaggerated tolerance \(\tau=0.01\). Report \(1\), with \((\ell_1,\mathbf d_1)=(2,(1,0))\), has rank \(2\); the more demanding report \(2\), with \((\ell_2,\mathbf d_2)=(3,(1,0))\), has rank \(3\). Suppose their exact thresholds are \(p^\star_1=5.000\) and \(p^\star_2=5.004\), in the correct order, but separate verified searches return \(p_{\mathrm{hi},1}=5.010\) and \(p_{\mathrm{hi},2}=5.004\). The uncorrected upper endpoints are reversed. Before final binary64 rounding, the corrected values are $G_1=5.010+0.01(2)=5.030$, $G_2=5.004+0.01(3)=5.034.$
The more demanding report again has the weakly higher critical value. The experiments use \(\tau=10^{-6}\); the larger value here only makes the correction visible.

It remains to compare utilities~\cite{myerson1981optimal,blumrosen2007combinatorial}.
For this proof only, write the critical value as \(+\infty\) when no permitted value is accepted. If the true duration and demand are infeasible, truthful reporting is rejected and gives zero utility. Every covering report remains infeasible, whereas any feasible report must understate a binding duration or demand and therefore gives no service value and utility at most zero. Truth is optimal in this case.
Now suppose the true duration and demand are feasible. Let \(q_i\) be the critical value for reporting them truthfully. If \(v_i\ge q_i\), truthful reporting is accepted and gives nonnegative utility \(v_i-q_i\); if \(v_i<q_i\), it is rejected and gives zero. A report that understates duration or demand does not cover the true task. If accepted, it gives utility \(-p\le0\); if rejected, it gives zero. Any covering report has duration and demand at least as large as the true task, so its finite critical value \(\hat q\) satisfies \(\hat q\ge q_i\), or it accepts no permitted value. When truth is accepted, such a report gives at most \(v_i-\hat q\le v_i-q_i\). When truth is rejected, \(v_i<q_i\le\hat q\), so the covering report is either rejected or accepted at negative utility. No deviation is profitable, and truthful utility is always nonnegative.

\section{Experiments}
\label{sec:experiments}

This section compares SP-SMM with other networks. We first describe training and imitation (\cref{sec:setup}) and the allocation streams (\cref{sec:benchmarks}), then the results (\cref{sec:main-result}). 

\subsection{Experimental setup}
\label{sec:setup}
\label{sec:train}
\label{sec:copy-method}

For RL, we run \cref{alg:sarsa} for \(8000\) episodes with five seeds and evaluate on \(200\) episodes.
\textsc{LinearMono} is the closest published baseline~\cite{stein2020sprl}. We take \textsc{LipschitzMono} to be a slope-bounded MLP~\cite{nolte2023expressive}, \textsc{SMNN} to combine monotone unit types~\cite{kim2024smnn}, \textsc{MonoDense} to stack constrained layers~\cite{runje2023constrained}, and \textsc{MonoNet} to use positive-weight layers~\cite{nguyen2019mononet}. We take \textsc{PMLinear} to be a sign-constrained linear layer~\cite{yanagisawa2022hlattice}. Neural models use Adam~\cite{kingma2015adam} and \textsc{LinearMono} uses projected SGD~\cite{stein2020sprl}. We use the verified search of \cref{alg:price-first}; the corresponding calculations for the other monotone networks are in \cref{app:invert-all}. Optimiser schedules, network sizes, inversion bit-widths and imitation budgets are in \cref{app:hparams}.

For imitation, we fit the same monotone network from teacher labels \(y(s)\in\{0,1\}\). We train the decision boundary with \(\mathcal{L}(w)=-y\log\sigma\bigl(\Delta_w(s)/\kappa\bigr)-(1-y)\log\bigl(1-\sigma\bigl(\Delta_w(s)/\kappa\bigr)\bigr)\), where \(\sigma\) is the logistic function and \(\kappa\) controls the margin scale. We then freeze that boundary and run fitted-SARSA sweeps on selected teacher transitions (\cref{eq:sarsa}), so the shared state-value term is fitted without moving the imitated sign of \(\Delta_w\). We do not follow this with RL episodes: imitation replaces the RL route rather than merely initialising it.
We separate the \emph{teacher}, which defines \(y(s)\), from the \emph{collection protocol}, which determines the states labelled. BC labels teacher-visited states~\cite{ross2010imitation}; DAgger labels and aggregates student-visited states~\cite{ross2011dagger}; DART labels states downstream of perturbed teacher actions~\cite{laskey2017dart}. We take DAggerW to use DAgger's data but to select the fitted round with highest held-out welfare rather than lowest \(\mathcal{L}\). The loss gives equal weight to errors with different welfare consequences; DAggerW addresses that mismatch at selection time.

We use two teachers. One is a hand-specified \textsc{Union} rule (\cref{sec:benchmarks}). The other is a receding-horizon planner. At each state we draw forecasts \(F^{(1)},\ldots,F^{(N)}\) from the arrival process and solve two mixed-integer packing problems per forecast, one forced to accept and one forced to reject the current task. Writing \(W_{\mathrm{pack}}(1,F)\) and \(W_{\mathrm{pack}}(0,F)\) for their welfare, we accept if and only if $ \sum_{n=1}^{N} W_{\mathrm{pack}}\bigl(1,F^{(n)}\bigr)
  \ge
  \sum_{n=1}^{N} W_{\mathrm{pack}}\bigl(0,F^{(n)}\bigr).$
The planner samples possible arrivals rather than reading the realised future. We do not use a clairvoyant mixed-integer teacher, because its labels could depend on future information absent from the student's inputs. 
\subsection{Benchmarks}
\label{sec:benchmarks}

\begin{table*}[t]
  \centering
  \caption{Allocation streams.}
  \label{tab:streams}
  \small
  \setlength{\tabcolsep}{4pt}
\begin{tabular}{@{}llrlp{0.42\textwidth}c@{}}
\toprule
stream & \(\mathbf{b}\) & \(T\) & class probabilities & tasks & \(\mathrm{OPT}\)\\
\midrule
\textsc{Partner} & \([3,1,1,3]\) & \(80\) &
  \((0.22,0.22,0.28,0.28)\) &
  A,B: \(\{r_1,r_2\}\), \(\{r_3,r_4\}\), \(v{=}24\), \(\ell{=}6\);
  C,D: \(r_1\)/\(r_4\), \(v{=}6\), \(\ell{=}12\) &
  LP\\
\textsc{Small} & \(10\) & \(20\) &
  fixed &
  \(v{=}1\), \(\ell{=}11\) except \(t{=}11\): \(v{=}15\), \(\ell{=}1\) &
  exact\\
\textsc{Large} & \(5\) & \(100\) &
  \(0.1/0.9\) &
  \(v{=}100\), \(\ell{=}1\); else \(v{=}20\), \(\ell{=}25\) &
  exact\\
\textsc{Cont.} & \([20,10]\) & \(80\) &
  \(0.15/0.85\) &
  \(\ell\sim\mathrm{Unif}\{1,\ldots,15\}\); high-density mixed demand, else sparse &
  LP\\
\textsc{Compl.} & \([6,6]\) & \(80\) &
  \(0.30/0.30/0.40\) &
  joint \([1,1]\): premium \(v{=}10\), \(\ell{=}2\) or long \(v{=}2.5\), \(\ell{=}10\);
  else \(v{=}1.5\), \(\ell{=}10\), \(3\) units of one resource &
  LP\\
\bottomrule
\end{tabular}

\end{table*}

\Cref{tab:streams} lists the five streams. In particular, Partner is the two-pair instance of \cref{ex:partner}, so one occupancy coefficient cannot serve both low-value classes. By contrast, \textsc{Small} and \textsc{Large} are the single-resource streams of~\citet{stein2020sprl}, on which a linear boundary is already competitive. \textsc{Continuous} mixes high-density multi-resource demand with a sparse class, and \textsc{Complementary} pairs single-resource fillers with joint-demand premium tasks. We report welfare as a fraction of an offline optimum \(\mathrm{OPT}\) from the complete arrival sequence: an LP upper bound where the table lists LP, and the integer packing optimum where it lists exact.
We compare against \textsc{LinearMono}~\cite{stein2020sprl} and the monotone networks of \cref{sec:setup}: \textsc{LipschitzMono}~\cite{nolte2023expressive}, \textsc{SMNN}~\cite{kim2024smnn}, \textsc{MonoDense}~\cite{runje2023constrained}, \textsc{MonoNet}~\cite{nguyen2019mononet}, and \textsc{PMLinear}~\cite{yanagisawa2022hlattice}. In addition, \textsc{HLattice} and CalibratedLattice (C-Lattice) are lattice baselines used here for cloning~\cite{you2017dln,yanagisawa2022hlattice}. U-MLP constrains none of value, duration and demand, whereas V-MLP is monotone only in reported value; neither is a strategyproof mechanism. Hand-specified references are first-come, first-served (FCFS) and \textsc{Union}, an in-sample rule that accepts a filler when its partner is busy, or its own resource is plentiful. Sampling, posted prices~\cite{stein2020sprl}, and the \textsc{LinearThresh} family are in \cref{app:benchmarks}. 

\subsection{Results}
\label{sec:main-result}
\label{sec:distil}

\begin{figure}[t]
  \centering
  \includegraphics[width=\columnwidth]{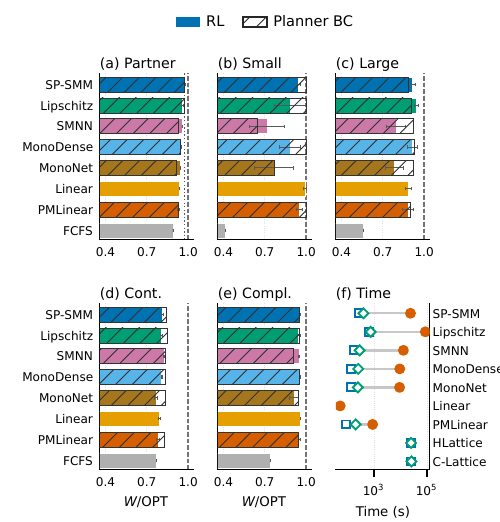}
  \Description{RL welfare over OPT on five streams with planner cloning as hatched bars, and Partner training time.}
  \caption{(a)--(e)~RL \(W/\mathrm{OPT}\) (mean \(\pm\) s.e.); hatched bars are planner cloning.
  (f)~Partner time: circles RL, squares Union BC, diamonds planner BC.}
  \label{fig:results}
\end{figure}

\begin{figure}[!htb]
  \centering
  \includegraphics[width=\columnwidth]{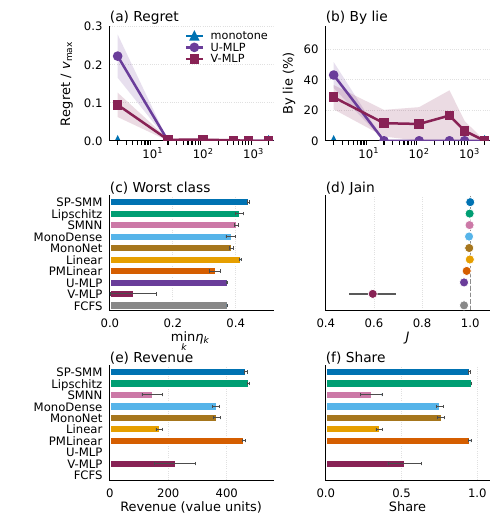}
  \Description{Partner-stream regret, by-lie rate, class service, Jain index, revenue and revenue share.}
  \caption{Partner stream.
  (a)--(b)~misreport search;
  (c)--(d)~class service;
  (e)--(f)~revenue in value units.}
  \label{fig:audit}
\end{figure}

On the partner stream (\cref{fig:results}), SP-SMM reaches \(0.978\) of \(\mathrm{OPT}\). In particular, it cuts the remaining gaps of \textsc{LinearMono} (\(6.2\%\)) and FCFS (\(11.0\%\)) by \(65\%\) and \(80\%\), respectively, and matches the in-sample \textsc{Union} rule (\(0.973\)).
By contrast, a linear boundary is already competitive when a single occupancy weight suffices: \textsc{LinearMono} reaches \(0.992\) on \textsc{Small} and, together with \textsc{MonoDense}, \(0.959\) on \textsc{Complementary}. On \textsc{Large}, \textsc{LipschitzMono} reaches \(0.946\), ahead of SP-SMM at \(0.912\). On \textsc{Continuous}, \textsc{SMNN} reaches \(0.824\) and SP-SMM \(0.812\). We fit separate weights for each stream. These comparisons therefore support a nonlinear rule precisely where two occupancy interactions cannot share coefficients. The corresponding numbers, with standard errors, are in \cref{app:main-results}.
Importantly, Partner welfare is close among the monotone networks, yet SP-SMM is the only one that is jointly first on welfare and worst-class service while still extracting more than \(0.94\) of welfare as critical-value revenue, as does \textsc{LipschitzMono} (\cref{fig:results,fig:audit}). \textsc{SMNN} and \textsc{LinearMono} extract about one third. Moreover, \textsc{LipschitzMono}'s RL run is several times slower, and a Lipschitz bound can restrict approximation~\cite{igel2024smm,nolte2023expressive}. Lattice cloning reaches similar Partner welfare, but its fitting time is of the same order as SP-SMM's RL run, in line with lattice parameter growth in the constrained coordinates~\cite{igel2024smm,you2017dln,yanagisawa2022hlattice}. In addition, SMM units give a value-axis derivative, so the verified search can try interval Newton (\cref{sec:invert}).

Consequently, imitation preserves almost all of this Partner welfare at a fraction of the RL cost (\cref{fig:results}(f)). Behaviour cloning of the \textsc{Union} teacher reaches \(0.975\) of \(\mathrm{OPT}\) in \(261\) seconds, compared with \(0.978\) in \(23{,}255\) seconds (\(6.46\) hours) for RL, so imitation is \(89.1\) times faster in these runs. Cloning the short-horizon planner reaches \(0.973\) in \(386\) seconds. Times include teacher rollouts, label fitting, SARSA value fitting and held-out checks for cloning; they exclude the cached rule search and exclude final evaluation for both. Both routes used the same SMM architecture, Partner stream, certified pricing path and cluster request (4 CPUs, 8\,GB). Variation among cluster nodes means this is a measured comparison, not an architecture-independent complexity result. \textsc{HLattice} and CalibratedLattice (C-Lattice) have cloning times only. No imitation protocol is best on every stream. The remaining architectures, protocols, and the Union-teacher search are in \cref{app:imitate}.

In addition, high Partner welfare comes with even class service and high payment (\cref{fig:audit}). Jain's index is \(1\) when every class is served at the same rate, not when every arrival is served~\cite{jain1984fairness}. SP-SMM serves about \(44\%\) of each class, so \(J=1.000\) while worst-class service is \(0.440\). By contrast, V-MLP starves a filler in four seeds, which lowers both \(J\) and the worst-class rate. The index formula and the remaining methods are in \cref{app:fair}.
U-MLP charges nothing, so it is omitted from \cref{fig:audit}(e)--(f).
At the most exploitable checkpoint, \(43\%\) of U-MLP's audited states and \(29\%\) of V-MLP's have a profitable lie. After training, U-MLP's zero regret coincides with charging \(0\), whereas SP-SMM and \textsc{LinearMono} remain at zero regret while still selecting. The misreport grid and remaining scores are in \cref{app:scratch}.

\section{Conclusions and Future Work}
\label{sec:conclusion}

Online resource allocation requires both adaptive decisions under uncertainty and truthful reporting by self-interested participants. To achieve this, SP-SMM combines three novel components. First, a monotonic neural network learns nonlinear occupancy interactions by construction. Second, a generalised critical payment charges the lowest accepted value with the other inputs fixed. Third, the same network can be trained by RL or by imitation. Together, monotonicity and that payment guarantee strategyproofness and individual rationality, while the network provides flexibility and imitation learning improves training efficiency. Importantly, those guarantees do not depend on the training procedure, since the monotonicity is architectural.
In particular, on a four-resource stream with two partner pairs, SP-SMM reaches \(97.8\%\) of the offline upper bound.
%By contrast, a linear rule remains competitive when one occupancy weight is enough. 
Moreover, imitation reaches \(97.5\%\) in 261 seconds, about 89 times faster than reinforcement learning. %However, no imitation protocol is best on every stream.

%Finally,
Two modelling choices remain open for future work. In the current work, we assumed observability of the arrival time of tasks. We plan to relax this assumption and use pre-commitment~\cite{stein2012precommit,stroehle2014online} to keep strategyproofness when owners can delay or otherwise misreport when they appear~\cite{gerding2011ev,robu2013jair}.
Moreover, we assumed that each task owner interacts once with the mechanism. We plan to extend this to iterative interactions using sequential mechanisms~\cite{sikdar2021sequential}, so that incentives cover a sequence of reports and associated payments.

\begin{acks}
The authors acknowledge support from the Engineering and Physical Sciences Research Council (EPSRC) through the Turing artificial intelligence (AI) Fellowship ``Citizen-Centric AI Systems (CCAIS)'' (EP/V022067/1) and FEVER project (EP/W005883/1), as well as the IRIDIS High Performance Computing Facility and the Low Carbon Comfort Centre (LCCC) at the University of Southampton.
\end{acks}

%%%%%%%%%%%%%%%%%%%%%%%%%%%%%%%%%%%%%%%%%%%%%%%%%%%%%%%%%%%%%%%%%%%%%%%%
%%% Appendices (former supplementary material, now included inline)
%%%%%%%%%%%%%%%%%%%%%%%%%%%%%%%%%%%%%%%%%%%%%%%%%%%%%%%%%%%%%%%%%%%%%%%%
\FloatBarrier
\bibliographystyle{ACM-Reference-Format} 
\bibliography{references}
\clearpage
\appendix
\counterwithin{table}{section}
\section{Critical payments for every monotone network}
\label{app:invert-all}

A verified critical payment requires two numerical guarantees: each network evaluation must enclose the exact output of the stored checkpoint, and the search must keep the first accepted value inside a proved rejecting--accepting bracket. The main text states this procedure as \cref{alg:price-first} in \cref{sec:invert}, together with the finite-precision correction in that section. We derive both guarantees from the underlying interval arithmetic before describing the architecture-specific calculations.

\subsection{Verified arithmetic and threshold search}
\label{app:cert-arithmetic}

We separate three intervals. An \emph{input interval} describes the values at which a network operation is evaluated. An \emph{output ball} encloses every exact result of that operation over the input interval. A \emph{root bracket} \([u_{\mathrm{lo}},u_{\mathrm{hi}}]\) records where the first accepted value can lie (\cref{sec:invert}). Neither an output ball nor a root bracket is a probability statement.
We represent a real number by the ball
\begin{equation}
  \label{eq:ball}
  B=[m_B-\rho_B,m_B+\rho_B],
  \qquad
  \rho_B\ge 0,
\end{equation}
with midpoint \(m_B\) and radius \(\rho_B\). We import each stored checkpoint parameter exactly into Arb and enclose the result of every subsequent operation~\cite{johansson2017arb}. For
\[
 B_1=[m_1-\rho_1,m_1+\rho_1],
 \qquad
 B_2=[m_2-\rho_2,m_2+\rho_2],
\]
the exact set of pairwise sums is
\begin{align}
 B_1+B_2
 &=
 \bigl[(m_1+m_2)-(\rho_1+\rho_2),\nonumber\\
 &\hspace{2.8em}(m_1+m_2)+(\rho_1+\rho_2)\bigr].
 \label{eq:ball-addition}
\end{align}
A convenient enclosure of all products has midpoint \(m_1m_2\) and radius
\begin{equation}
 |m_1|\rho_2+|m_2|\rho_1+\rho_1\rho_2.
 \label{eq:ball-product}
\end{equation}
These formulae describe exact real arithmetic. We must also round the stored midpoint and radius: Arb rounds the lower endpoint downward and the upper endpoint upward, enlarging the ball just enough to retain containment~\cite{johansson2017arb}.

For an increasing elementary function \(h\), the exact image of \([a,b]\) is \([h(a),h(b)]\). Arb evaluates and outward-rounds those endpoints, and its validated implementations similarly enclose the exponential, logarithm, and other transcendental operations used by softplus and the network head~\cite{johansson2017arb}. Induction over the computation graph therefore yields the invariant used in \cref{sec:invert}: the reported output ball contains the exact real result obtained from the stored checkpoint. Even a point input can produce a nonzero radius, because intermediate results must be rounded.

Interval evaluation can be conservative. Evaluating \(z-z\) while treating its two appearances as independent copies of \([a,b]\) gives \([a-b,b-a]\), although the true result is exactly zero. This \emph{dependency effect} can widen a ball but cannot exclude the exact answer~\cite{johansson2017arb}. Increasing the working precision reduces rounding error; subdividing an input interval can additionally reduce dependency overestimation.

Consequently, the sign tests themselves are those of \cref{sec:invert}. One additional endpoint convention is required by the tie rule: \(U(u)=0\) proves only \(F_s(u)\le 0\), whereas acceptance needs a nonnegative lower endpoint. For example, \([-0.03,-0.01]\) proves rejection, \([0.01,0.03]\) proves acceptance, and \([-0.01,0.02]\) proves neither.

Bisection of that rejecting--accepting bracket needs no slope estimate and cannot overshoot the root. Starting from \([0,1]\), after \(n\) successful steps
\[
 u_{\mathrm{hi}}-u_{\mathrm{lo}}\le2^{-n},
 \qquad
 p_{\mathrm{hi}}-p_{\mathrm{lo}}
 \le v_{\max}2^{-n}.
\]
After outward conversion we compare the two binary64 endpoints as exact dyadic fractions, so we verify the requested width without a further rounded subtraction. The method gains at most one binary digit per step, and a ball that straddles zero cannot move either endpoint; that case is arithmetic and is handled by the precision restart in \cref{alg:price-first} of \cref{sec:invert}.

\paragraph{Safe acceleration with interval Newton.}
Point Newton sends \(m\) to \(m-F_s(m)/F'_s(m)\). A rounded derivative, a nearly zero slope, or an unfortunate initial point can send the iterate outside the current bracket~\cite{nocedal2006numerical}.
We enclose a derivative interval \(F'_s(I)\) that contains \(F'_s(z)\) for every \(z\) in the current bracket \(I=[a,b]\)~\cite{johansson2017arb}. If this interval contains zero, interval division is unsafe and we discard the proposal; this includes every bracket that reaches the outer clamp. Otherwise, with \(m=(a+b)/2\), the interval-Newton image is
\begin{equation}
 \label{eq:interval-newton}
 N(I)=m-\frac{F_s(m)}{F'_s(I)}.
\end{equation}
Suppose \(u^\star\in I\) is a root. The mean-value theorem gives some \(\xi\) between \(m\) and \(u^\star\) such that
\[
 0=F_s(u^\star)
  =F_s(m)+F'_s(\xi)(u^\star-m).
\]
Rearranging places \(u^\star\) at \(m-F_s(m)/F'_s(\xi)\). Since \(F'_s(\xi)\in F'_s(I)\), validated interval division proves \(u^\star\in N(I)\). We intersect \(N(I)\) with \(I\). A proposal becomes the new bracket only after the endpoint tests of \cref{sec:invert} restore the rejecting--accepting invariant.

\subsection{\textsc{LinearMono}}
The same search applies to each monotone architecture. For \textsc{LinearMono} we represent
\[
\Delta_w(\phi(s,u))=(\mathbf{w}^{\mathrm{acc}}-\mathbf{w}^{\mathrm{rej}})\cdot\phi(s,u).
\]
We write \(j_v\) for the normalised-value coordinate and \(w_u=w^{\mathrm{acc}}_{j_v}-w^{\mathrm{rej}}_{j_v}\). If \(w_u=0\), \(\Delta_w\) is constant along the value axis, so the endpoint test of \cref{sec:invert} either accepts every permitted value or none. If \(w_u>0\), we take the mathematical threshold to be
\begin{equation}
  \label{eq:lin-cv}
  u^\star
    \;=\;
  \max\Biggl(0,\;
    \frac{\sum_{j\neq j_v}\phi_j\,(w^{\mathrm{rej}}_j-w^{\mathrm{acc}}_j)}{w_u}
  \Biggr).
\end{equation}
If the untruncated fraction exceeds \(1\), every permitted value is rejected. We keep \(w_u\ge0\) by the coefficient projection of~\citet{stein2020sprl}.

We import the stored binary coefficients into Arb and perform that division with outward rounding~\cite{johansson2017arb}. This expression is the mathematical threshold \(u^\star\); the deployed critical value remains the finite-precision correction of \cref{sec:invert}.

\paragraph{Example.}
We take \(v_{\max}=24\) and \(\Delta_w=2u-0.5\). Then \(u^\star=0.25\) and \(p^\star=6\).

\subsection{Hard min--max model}
By contrast, we obtain Sill's hard min--max network by replacing the smooth extrema of \cref{sec:smm} with ordinary maxima and minima~\cite{sill1997monotonic}. Because every value-axis weight is positive (\cref{sec:smm}), we take the hard threshold for a target level \(L\) to be
\begin{equation}
  \label{eq:sill}
  u_{\mathrm{hard}}
  =
  \max_{k\le K}\min_{h\le H}
  \frac{L-a_{kh}}{w_{kh}}.
\end{equation}
A unit \(h\) reaches \(L\) at \((L-a_{kh})/w_{kh}\). The group maximum reaches \(L\) when its earliest unit does, hence the inner minimum; the network minimum reaches \(L\) only when every group does, hence the outer maximum.

The implemented SMM uses log-sum-exp (\cref{sec:smm}), so every affine unit contributes and \cref{eq:sill} is not the smooth root. We therefore do not seed the search with \(u_{\mathrm{hard}}\). The formula is included only to explain why the hard network has a closed form while SMM does not.

\subsection{Smooth min-max model}
Instead, we take the value-axis derivative of a smooth maximum to be a softmax-weighted average of its unit slopes~\cite{igel2024smm}. Writing
\[
  \pi_{kh}
  =
  \frac{\exp(\zeta_{kh})}{\sum_{h'}\exp(\zeta_{kh'})},
  \qquad
  \alpha_k
  =
  \frac{\exp(-\operatorname{smax}_h\zeta_{kh})}
       {\sum_{k'}\exp(-\operatorname{smax}_h\zeta_{k'h})},
\]
we obtain, before the output head,
\[
  f'_w(u)
  =
  \mathrm e^{-\beta}
  \sum_{k\le K}\alpha_k
  \sum_{h\le H}\pi_{kh}w_{kh}
  >0.
\]
The context term of \cref{sec:smm} does not depend on \(u\). Each map of the bounded difference in \cref{sec:smm} is non-decreasing, so the chain rule multiplies \(f'_w\) by a nonnegative factor; that factor is zero on the outer clamp. We therefore enclose the complete derivative of \(F_s\) and apply interval Newton only as in \cref{app:cert-arithmetic}.

\subsection{Critical values}
The remaining monotone networks have no closed form. We invert them by bisection of \(F_s\) with verified signs (\cref{sec:invert}), not by a closed-form root. None of \textsc{LipschitzMono}, \textsc{SMNN}, \textsc{MonoDense} or \textsc{MonoNet} has an algebraic inverse.

A lattice is piecewise linear inside each cell~\cite{you2017dln}. Neither a sign-constrained linear map nor a lattice makes the complete action-value difference piecewise linear, because the bounded difference in \cref{sec:smm} applies a nonlinear \(\tanh\) head. We therefore bisect the full \(F_s\) for \textsc{PMLinear} as well, rather than interpolating an inner map whose sign is not the acceptance test.
U-MLP and V-MLP are controls and are not run through this inversion (\cref{app:scratch}).

\section{Notation}
\label{app:notation}

\Cref{tab:notation-full} shows the principal symbols used in the main text. 

\begin{table}[t]
  \centering
  \caption{Principal notation used in the main text.}
  \label{tab:notation-full}
  \small
  \begin{tabular}{@{}p{0.30\columnwidth}p{0.62\columnwidth}@{}}
\toprule
Symbol & Meaning \\
\midrule
$m,r,b_r,T,t,i$ & number of resource types; resource index; capacity; horizon; slot; arrival \\
$\theta_i=(v_i,\ell_i,\mathbf{d}_i)$ & true type \\
$\hat\theta_i$, $\hat\Theta_{<i}$ & report; history before $i$ \\
$s$, $\mathbf f(s)$, $f_r(s)$ & state; free-capacity vector and its resource-\(r\) coordinate \\
$\mathcal{M}=(x,p)$ & allocation and payment \\
$W$, $U_i$, $c$, $\mathbf{1}[\cdot]$ & welfare; utility of $i$; covering indicator; indicator \\
$R$, $J_t$, $Q_w$, $\Delta_w$, $w$, $\pi_w$ & reward; return from \(t\); action values; accept--reject action-value difference; parameters; greedy policy \\
$v_{\max}$, $\ell_{\max}$, $o(s)$, $\phi(s)$, $z(s)$ & bounds; occupancy; features; context \\
$\mathbf{w}^{\mathrm{acc}},\mathbf{w}^{\mathrm{rej}}$ & LinearMono accept and reject weights \\
$u$, $u^\star$, $\phi(s,u)$, $F_s(u)$ & normalised candidate value; threshold; features at \(u\); value-axis action-value difference \\
$K,H,k,h$ & numbers of SMM groups and units; their indices \\
$a_{kh},w_{kh},\beta,\gamma$ & SMM intercepts, slopes, scale, offset \\
$f_{\mathrm{SMM}}$, $f_{\mathrm{ctx}}$, $f_w$ & SMM core; context MLP; $f_{\mathrm{SMM}}+f_{\mathrm{ctx}}$ \\
$\mu,\nu,\lambda,\lambda_{\max},b_0$ & output-head centre, scale, gain, cap, and offset \\
$B,m_B,\rho_B$ & arithmetic ball; midpoint; radius \\
$I=[a,b]$, $m$, $F'_s(I)$, $N(I)$, $\xi$ & current interval; midpoint; derivative enclosure; interval-Newton image; intermediate point \\
$u_{\mathrm{lo}},u_{\mathrm{hi}}$, $\tau$, $P$ & root bracket; critical-value tolerance; working precision \\
$p^\star$, $p_{\mathrm{lo}}$, $p_{\mathrm{hi}}$, $q(s)$, $p(s)$ & mathematical threshold; lower and upper enclosure endpoints; deployed critical value; payment charged \\
$\operatorname{rank}(\hat\ell,\hat{\mathbf d})$ & integer report rank used by the finite-precision correction \\
$\lceil\cdot\rceil_{\mathbb F}$, $\rho_q$, $\operatorname{ulp}$ & upward binary64 rounding; its increase; unit in the last place \\
$\varepsilon$, $\overline{W}_{\mathrm{FCFS}}$, $s',a'$ & exploration; FCFS scale; SARSA successor \\
$y,\kappa,N,F^{(n)},W_{\mathrm{pack}}$ & teacher label; margin scale; number of forecasts; packing welfare \\
$\sigma(\chi)$ & logistic sigmoid $(1+\mathrm{e}^{-\chi})^{-1}$ \\
$\alpha_r,\tau_c,x^{\mathrm{LT}}$ & LinearThresh weights, class threshold, allocation \\
$r(c),r^\star(c),\delta_c$ & filler resource, partner, Union threshold \\
$\bar p$, $\eta_k$, $J(\boldsymbol{\eta})$ & posted price; class service rate; Jain index \\
\bottomrule
\end{tabular}

\end{table}

\section{Allocation streams and reference rules}
\label{app:benchmarks}

The five streams, including Partner as \cref{ex:partner} in \cref{sec:mdp}, are listed in \cref{sec:benchmarks}. This section presents sampling, the offline programmes, and the hand-specified rules that the main text names. In particular, fillers occupy capacity for twice as long as premium tasks while providing one-quarter of the value, and offered load (expected requested resource-time divided by available resource-time) is \(1.32\)--\(1.56\).

On \textsc{Continuous}, the high-density class in \cref{sec:benchmarks} draws each demand uniformly from the positive integers up to \([5,3]\) and value density from \(\mathcal N(5,0.5^2)\), truncated below at \(0.01\); otherwise each demand may be zero (an all-zero draw becomes one unit of the first resource) and density from \(\mathcal N(1,0.3^2)\), again truncated below at \(0.01\). Value is that density times \(\ell\sum_{r=1}^{m}d_r\), capped at \(v_{\max}\). On \textsc{Complementary}, single-resource tasks create unequal free capacities, while premium and long-duration joint-demand tasks need the two resources together.

\label{sec:opt}
Where that table lists LP, \(\mathrm{OPT}\) is the LP relaxation of the integer packing problem, one variable per task and one capacity constraint per resource and time, solved with SciPy's HiGHS interface~\cite{virtanen2020scipy,huangfu2018highs}. The relaxation may accept fractions of tasks, so \(W/\mathrm{OPT}\) is then a ratio to an upper bound. Where the table lists exact, \(\mathrm{OPT}\) is the integer offline optimum obtained by min-cost flow~\cite{hagberg2008networkx}. A ratio of \(1\) means the online rule matched this programme.

In addition, \textsc{LinearThresh} and \textsc{Union} are hand-specified families, not trained learners. For \textsc{LinearThresh}, let \(\alpha_r\) be a global weight on an idle unit of resource \(r\) and \(\tau_c\) a cutoff for class \(c\):
\(x^{\mathrm{LT}}(s)=1\) if and only if \(\sum_{r=1}^{m}\alpha_r f_r(s)\ge\tau_c\).
Every class uses the same \(\boldsymbol{\alpha}=(\alpha_1,\ldots,\alpha_m)\). Fillers with different scarce partners need different class--occupancy interactions, which one global weight vector cannot express. \textsc{LinearMono} additionally reads time and the remaining busy time of every capacity unit; however, it still uses one occupancy coefficient per resource, so the four decisions of \cref{ex:partner} in \cref{sec:mdp} cannot hold together.

\textsc{Union} accepts every feasible multi-resource task. For a class \(c\) that uses only resource \(r(c)\), it looks for a two-resource class that also uses \(r(c)\); if one exists, the other resource of that class is the partner \(r^\star(c)\). With class threshold \(\delta_c\), the one-resource task is accepted when \(f_{r^\star(c)}(s)=0\) or \(f_{r(c)}(s)\ge\delta_c\). If no partner is found, only the capacity-threshold condition applies, so the rule reduces to per-class capacity thresholds on a single-resource stream.

For each family we report the parameter choice with highest welfare on a shared reference block of \(200\) episodes (seeds \(10000\)--\(10199\)) from the streams in \cref{sec:benchmarks}. These are deliberately favourable, in-sample values showing what each hand-specified family can express; they are not held-out estimates for a trained learner. We also include the posted-price rule of~\citet{stein2020sprl}. That rule uses one occupancy-blind price \(\bar p\ge 0\): a feasible report is accepted iff \(\hat v\ge \bar p\cdot\hat\ell\sum_{r=1}^{m}\hat d_r\). We search \(\bar p\in[0,4]\) on the reference block. On the partner stream every class occupies \(12\) unit-slots, so a single \(\bar p\) cannot separate fillers from premium tasks.

\section{Training and inversion settings}
\label{app:hparams}

In addition to the episode counts, seeds, and imitation loss of \cref{sec:setup}, \cref{tab:hparams} records optimiser schedules, network sizes, inversion bit-widths, and imitation budgets.

\begin{table}
  \centering
  \caption{Training and imitation settings.}
  \label{tab:hparams}
  \small
  \begin{tabular}{@{}p{0.56\columnwidth}p{0.38\columnwidth}@{}}
\toprule
quantity & value\\
\midrule
$\varepsilon$ & $0.1\to 0$ linearly\\
\textsc{LinearMono} & SGD, learning rate $0.1\to 0$\\
SP-SMM / SMNN & Adam~\cite{kingma2015adam}, $5\cdot10^{-3}\to10^{-5}$\\
other neural models & Adam, $10^{-2}\to10^{-5}$\\
gradient clipping (neural) & $\ell_2$ norm at $5$\\
fitted SARSA refresh / passes / minibatch & $20$ episodes / $5$ / $64$\\
SARSA discount & $1$ (undiscounted)\\
reward scale & $1/\overline{W}_{\mathrm{FCFS}}$ ($20$ FCFS episodes)\\
neural TD targets & clipped to $[-5,5]$\\
gain cap $\lambda_{\max}$ & $2$\\
SMM $K\times H$ & $6\times 6$\\
context MLP width & $64$ (sigmoid hidden layer, linear output)\\
DART action-flip probability & $0.1$~\cite{laskey2017dart}\\
reported experiments & verified interval upper endpoint\\
complete mechanism & verified interval, then finite-precision correction\\
critical-value tolerance / precision & $10^{-6}$ / $64,128,256$ bits\\
teacher-transition / labelled episodes & $200$ / $50$\\
imitation \(\mathcal{L}\) passes / minibatch / $\kappa$ & $30$ / $64$ / $0.05$\\
value fitting sweeps / passes per sweep & $20$ / $5$\\
DAgger rounds / selection rollouts & $5$ / $10$\\
planner forecasts / look-ahead slots & $3$ / $24$\\
\bottomrule
\end{tabular}

\end{table}

\section{Numeric reinforcement learning welfare}
\label{app:main-results}

Using those settings, \cref{tab:main} is the numeric counterpart of the welfare figure in \cref{sec:main-result}: it adds standard errors, Partner RL training time, and the in-sample \textsc{LinearThresh} and posted-price rows that the figure omits.
\begin{table*}[t]
  \centering
  \caption{RL welfare divided by the offline optimum, with Partner training time.}
  \label{tab:main}
  \small
  \begin{tabular}{@{}lccccccc@{}}
\toprule
& \multicolumn{5}{c}{welfare / $\mathrm{OPT}$} & \multicolumn{2}{c}{\textsc{Partner}}\\
\cmidrule(lr){2-6}\cmidrule(l){7-8}
method & \textsc{Partner} & \textsc{Small} & \textsc{Large} & \textsc{Cont.} & \textsc{Compl.} & gap & train (s)\\
\midrule
SP-SMM & $\mathbf{0.978\pm0.001}$ & $0.942\pm0.017$ & $0.912\pm0.021$ & $0.812\pm0.010$ & $0.953\pm0.003$ & $\mathbf{2.2\%}$ & 23,255\\
\textsc{LipschitzMono}~\cite{nolte2023expressive} & $0.960\pm0.006$ & $0.883\pm0.117$ & $\mathbf{0.946\pm0.014}$ & $0.804\pm0.007$ & $0.946\pm0.010$ & 4.0\% & 84,607\\
\textsc{SMNN}~\cite{kim2024smnn} & $0.957\pm0.001$ & $0.717\pm0.127$ & $0.795\pm0.072$ & $\mathbf{0.824\pm0.006}$ & $0.949\pm0.006$ & 4.3\% & 12,558\\
\textsc{MonoDense}~\cite{runje2023constrained} & $0.946\pm0.003$ & $0.883\pm0.076$ & $0.914\pm0.037$ & $0.808\pm0.003$ & $0.959\pm0.001$ & 5.4\% & 9,119\\
\textsc{MonoNet}~\cite{nguyen2019mononet} & $0.943\pm0.001$ & $0.767\pm0.143$ & $0.785\pm0.066$ & $0.769\pm0.011$ & $0.913\pm0.031$ & 5.7\% & 9,003\\
\textsc{LinearMono}~\cite{stein2020sprl} & $0.938\pm0.002$ & $\mathbf{0.992\pm0.008}$ & $0.888\pm0.023$ & $0.792\pm0.007$ & $\mathbf{0.959\pm0.000}$ & 6.2\% & 51\\
\textsc{PMLinear}~\cite{yanagisawa2022hlattice} & $0.929\pm0.005$ & $0.950\pm0.020$ & $0.882\pm0.043$ & $0.781\pm0.009$ & $0.951\pm0.004$ & 7.1\% & 861\\
\midrule
\multicolumn{8}{@{}l}{\textit{Reference policies (fixed in-sample episode block)}}\\
FCFS & 0.890 & 0.417 & 0.559 & 0.770 & 0.741 & 11.0\% & --\\
posted-price & 0.890 & 0.625 & 0.764 & 0.821 & 0.954 & 11.0\% & --\\
\textsc{LinearThresh} & 0.917 & 1.000 & 0.960 & 0.795 & 0.956 & 8.3\% & --\\
\textsc{Union} & 0.973 & 1.000 & 0.960 & 0.791 & 0.936 & 2.7\% & --\\
\bottomrule
\end{tabular}

\end{table*}

\section{Imitation grid}
\label{app:imitate}
\label{app:imit-method}

Turning from RL welfare to imitation, the loss, protocols, and teachers are those of \cref{sec:setup}. In particular, DART's flip probability, the planner window, \(\kappa\), and the label budget appear in \cref{tab:hparams}. For the partner stream we compare the hand-specified families of \cref{app:benchmarks} on \(40\) episodes disjoint from training and final evaluation, and we take the selected \textsc{Union} rule as the teacher.

We then report, for every neural architecture, BC from that rule and planner BC, DAgger, DAggerW, and DART, using five seeds. Monotone rows use \cref{alg:price-first} in \cref{sec:invert}. \Cref{tab:imitate-all} reports \(W/\mathrm{OPT}\); \cref{tab:imitate-all-time} reports fitting time. No protocol ranks first on every stream. On \textsc{Large}, DAggerW exceeds DAgger for \textsc{LipschitzMono} (\(0.963\) versus \(0.934\)), as it does for SP-SMM in \cref{sec:main-result}; the same comparison changes little on the partner stream.

On the partner stream, U-MLP and V-MLP both stay near FCFS welfare (\(0.891\)) under most protocols, so dropping duration and demand monotonicity does not buy a welfare gain here. Neither a report constraint nor a strong teacher alone ensures high welfare; the teacher, collection protocol, and function class also matter. Class-level service after imitation is \cref{tab:imitate-fair}. V-MLP matches FCFS on that table under both protocols, so we omit its row.

\begin{table}
  \centering
  \caption{Imitation welfare divided by offline optimum.}
  \label{tab:imitate-all}
  \small
  \setlength{\tabcolsep}{2pt}
  \resizebox{\columnwidth}{!}{\begin{tabular}{@{}llccccc@{}}
\toprule
setting & method & Rule BC & Planner BC & DAgger & DAggerW & DART\\
\midrule
\textsc{Partner} & SP-SMM & $\mathbf{0.975\pm0.000}$ & $0.973\pm0.001$ & $0.972\pm0.002$ & $0.973\pm0.001$ & $0.971\pm0.002$\\
 & \textsc{LipschitzMono} & $\mathbf{0.975\pm0.001}$ & $0.971\pm0.001$ & $0.974\pm0.002$ & $0.974\pm0.002$ & $0.970\pm0.002$\\
 & \textsc{SMNN} & $0.937\pm0.002$ & $0.933\pm0.006$ & $0.933\pm0.002$ & $\mathbf{0.938\pm0.004}$ & $0.919\pm0.012$\\
 & \textsc{MonoDense} & $\mathbf{0.947\pm0.008}$ & $0.941\pm0.004$ & $0.940\pm0.003$ & $0.940\pm0.006$ & $0.937\pm0.006$\\
 & \textsc{MonoNet} & $\mathbf{0.938\pm0.002}$ & $0.917\pm0.003$ & $0.918\pm0.003$ & $0.931\pm0.002$ & $0.927\pm0.003$\\
 & \textsc{PMLinear} & $\mathbf{0.936\pm0.002}$ & $0.930\pm0.004$ & $0.927\pm0.002$ & $0.931\pm0.001$ & $0.928\pm0.003$\\
 & V-MLP & $\mathbf{0.891\pm0.000}$ & $\mathbf{0.891\pm0.000}$ & $\mathbf{0.891\pm0.000}$ & $\mathbf{0.891\pm0.000}$ & $\mathbf{0.891\pm0.000}$\\
 & U-MLP & $0.891\pm0.000$ & $0.891\pm0.000$ & $\mathbf{0.905\pm0.015}$ & $\mathbf{0.905\pm0.015}$ & $0.891\pm0.000$\\
\midrule
\textsc{Small} & SP-SMM & $\mathbf{1.000\pm0.000}$ & $\mathbf{1.000\pm0.000}$ & $\mathbf{1.000\pm0.000}$ & $\mathbf{1.000\pm0.000}$ & $\mathbf{1.000\pm0.000}$\\
 & \textsc{LipschitzMono} & $\mathbf{1.000\pm0.000}$ & $\mathbf{1.000\pm0.000}$ & $\mathbf{1.000\pm0.000}$ & $\mathbf{1.000\pm0.000}$ & $\mathbf{1.000\pm0.000}$\\
 & \textsc{SMNN} & $0.650\pm0.143$ & $0.650\pm0.143$ & $\mathbf{1.000\pm0.000}$ & $\mathbf{1.000\pm0.000}$ & $0.650\pm0.143$\\
 & \textsc{MonoDense} & $\mathbf{1.000\pm0.000}$ & $\mathbf{1.000\pm0.000}$ & $\mathbf{1.000\pm0.000}$ & $\mathbf{1.000\pm0.000}$ & $0.767\pm0.143$\\
 & \textsc{MonoNet} & $0.767\pm0.143$ & $0.767\pm0.143$ & $\mathbf{1.000\pm0.000}$ & $\mathbf{1.000\pm0.000}$ & $0.883\pm0.117$\\
 & \textsc{PMLinear} & $\mathbf{1.000\pm0.000}$ & $\mathbf{1.000\pm0.000}$ & $\mathbf{1.000\pm0.000}$ & $\mathbf{1.000\pm0.000}$ & $\mathbf{1.000\pm0.000}$\\
 & V-MLP & $0.417\pm0.000$ & $0.417\pm0.000$ & $0.417\pm0.000$ & $\mathbf{0.458\pm0.042}$ & $0.417\pm0.000$\\
 & U-MLP & $0.417\pm0.000$ & $0.417\pm0.000$ & $\mathbf{0.650\pm0.143}$ & $\mathbf{0.650\pm0.143}$ & $0.417\pm0.000$\\
\midrule
\textsc{Large} & SP-SMM & $\mathbf{0.960\pm0.000}$ & $0.887\pm0.026$ & $0.920\pm0.015$ & $0.945\pm0.009$ & $0.911\pm0.016$\\
 & \textsc{LipschitzMono} & $0.955\pm0.002$ & $0.911\pm0.021$ & $0.934\pm0.019$ & $\mathbf{0.963\pm0.004}$ & $0.915\pm0.015$\\
 & \textsc{SMNN} & $0.960\pm0.000$ & $0.926\pm0.024$ & $0.955\pm0.013$ & $\mathbf{0.963\pm0.008}$ & $0.929\pm0.012$\\
 & \textsc{MonoDense} & $0.960\pm0.000$ & $0.926\pm0.017$ & $0.939\pm0.010$ & $\mathbf{0.960\pm0.002}$ & $0.947\pm0.014$\\
 & \textsc{MonoNet} & $\mathbf{0.960\pm0.000}$ & $0.922\pm0.018$ & $0.894\pm0.028$ & $0.957\pm0.009$ & $0.935\pm0.019$\\
 & \textsc{PMLinear} & $\mathbf{0.960\pm0.000}$ & $0.898\pm0.016$ & $0.900\pm0.026$ & $0.931\pm0.010$ & $0.926\pm0.013$\\
 & V-MLP & $0.765\pm0.000$ & $0.783\pm0.019$ & $0.782\pm0.092$ & $\mathbf{0.905\pm0.037}$ & $0.765\pm0.000$\\
 & U-MLP & $0.345\pm0.213$ & $0.180\pm0.180$ & $\mathbf{0.641\pm0.082}$ & $\mathbf{0.641\pm0.082}$ & $0.181\pm0.181$\\
\midrule
\textsc{Cont.} & SP-SMM & $0.805\pm0.002$ & $0.841\pm0.006$ & $0.839\pm0.006$ & $\mathbf{0.848\pm0.005}$ & $0.841\pm0.003$\\
 & \textsc{LipschitzMono} & $0.787\pm0.006$ & $\mathbf{0.852\pm0.007}$ & $0.839\pm0.007$ & $0.849\pm0.004$ & $0.837\pm0.006$\\
 & \textsc{SMNN} & $0.795\pm0.005$ & $0.834\pm0.010$ & $0.812\pm0.011$ & $\mathbf{0.840\pm0.006}$ & $0.823\pm0.009$\\
 & \textsc{MonoDense} & $0.804\pm0.002$ & $0.838\pm0.006$ & $0.845\pm0.007$ & $\mathbf{0.852\pm0.002}$ & $0.834\pm0.006$\\
 & \textsc{MonoNet} & $0.806\pm0.002$ & $\mathbf{0.837\pm0.005}$ & $0.817\pm0.005$ & $0.828\pm0.002$ & $0.821\pm0.003$\\
 & \textsc{PMLinear} & $0.801\pm0.003$ & $0.828\pm0.005$ & $0.815\pm0.003$ & $\mathbf{0.835\pm0.005}$ & $0.825\pm0.003$\\
 & V-MLP & $0.769\pm0.001$ & $\mathbf{0.770\pm0.000}$ & $\mathbf{0.770\pm0.000}$ & $\mathbf{0.770\pm0.000}$ & $\mathbf{0.770\pm0.000}$\\
 & U-MLP & $\mathbf{0.770\pm0.000}$ & $\mathbf{0.770\pm0.000}$ & $\mathbf{0.770\pm0.000}$ & $\mathbf{0.770\pm0.000}$ & $\mathbf{0.770\pm0.000}$\\
\midrule
\textsc{Compl.} & SP-SMM & $0.956\pm0.000$ & $0.949\pm0.006$ & $0.952\pm0.003$ & $0.957\pm0.001$ & $\mathbf{0.958\pm0.000}$\\
 & \textsc{LipschitzMono} & $0.956\pm0.000$ & $0.951\pm0.005$ & $0.955\pm0.002$ & $0.956\pm0.002$ & $\mathbf{0.958\pm0.001}$\\
 & \textsc{SMNN} & $\mathbf{0.956\pm0.000}$ & $0.910\pm0.029$ & $0.925\pm0.021$ & $0.940\pm0.009$ & $0.941\pm0.009$\\
 & \textsc{MonoDense} & $0.956\pm0.000$ & $0.950\pm0.002$ & $0.954\pm0.003$ & $0.955\pm0.003$ & $\mathbf{0.957\pm0.001}$\\
 & \textsc{MonoNet} & $\mathbf{0.956\pm0.000}$ & $0.943\pm0.006$ & $0.951\pm0.004$ & $0.955\pm0.001$ & $0.953\pm0.004$\\
 & \textsc{PMLinear} & $\mathbf{0.956\pm0.000}$ & $0.947\pm0.004$ & $0.948\pm0.006$ & $0.948\pm0.006$ & $0.955\pm0.001$\\
 & V-MLP & $\mathbf{0.954\pm0.000}$ & $\mathbf{0.954\pm0.000}$ & $\mathbf{0.954\pm0.000}$ & $\mathbf{0.954\pm0.000}$ & $\mathbf{0.954\pm0.000}$\\
 & U-MLP & $\mathbf{0.785\pm0.043}$ & $0.742\pm0.000$ & $0.742\pm0.000$ & $0.742\pm0.000$ & $0.742\pm0.000$\\
\bottomrule
\end{tabular}
}
\end{table}

\begin{table}
  \centering
  \caption{Imitation fitting time in seconds.}
  \label{tab:imitate-all-time}
  \small
  \setlength{\tabcolsep}{2pt}
  \resizebox{\columnwidth}{!}{\begin{tabular}{@{}llccccc@{}}
\toprule
setting & method & Rule BC & Planner BC & DAgger & DAggerW & DART\\
\midrule
\textsc{Partner} & SP-SMM & 261 & 386 & 410 & 412 & 382\\
 & \textsc{LipschitzMono} & 643 & 724 & 751 & 739 & 808\\
 & \textsc{SMNN} & 173 & 278 & 339 & 472 & 268\\
 & \textsc{MonoDense} & 134 & 247 & 256 & 272 & 244\\
 & \textsc{MonoNet} & 133 & 237 & 273 & 292 & 279\\
 & \textsc{PMLinear} & 88 & 197 & 206 & 214 & 198\\
 & V-MLP & 109 & 219 & 221 & 220 & 217\\
 & U-MLP & 90 & 185 & 189 & 189 & 195\\
\midrule
\textsc{Small} & SP-SMM & 54 & 119 & 52 & 89 & 47\\
 & \textsc{LipschitzMono} & 64 & 71 & 71 & 76 & 63\\
 & \textsc{SMNN} & 34 & 38 & 37 & 37 & 38\\
 & \textsc{MonoDense} & 36 & 94 & 31 & 35 & 36\\
 & \textsc{MonoNet} & 28 & 26 & 27 & 29 & 26\\
 & \textsc{PMLinear} & 31 & 59 & 24 & 27 & 22\\
 & V-MLP & 37 & 37 & 27 & 33 & 27\\
 & U-MLP & 20 & 19 & 21 & 21 & 26\\
\midrule
\textsc{Large} & SP-SMM & 320 & 377 & 386 & 400 & 444\\
 & \textsc{LipschitzMono} & 961 & 879 & 915 & 978 & 884\\
 & \textsc{SMNN} & 200 & 267 & 367 & 528 & 269\\
 & \textsc{MonoDense} & 190 & 234 & 368 & 248 & 231\\
 & \textsc{MonoNet} & 160 & 227 & 223 & 219 & 203\\
 & \textsc{PMLinear} & 115 & 146 & 157 & 147 & 143\\
 & V-MLP & 142 & 168 & 166 & 173 & 225\\
 & U-MLP & 118 & 128 & 127 & 132 & 134\\
\midrule
\textsc{Cont.} & SP-SMM & 631 & 682 & 799 & 810 & 685\\
 & \textsc{LipschitzMono} & 952 & 980 & 952 & 1,014 & 890\\
 & \textsc{SMNN} & 309 & 500 & 682 & 572 & 486\\
 & \textsc{MonoDense} & 287 & 412 & 450 & 460 & 410\\
 & \textsc{MonoNet} & 211 & 420 & 451 & 500 & 412\\
 & \textsc{PMLinear} & 101 & 344 & 401 & 367 & 347\\
 & V-MLP & 129 & 363 & 369 & 375 & 363\\
 & U-MLP & 90 & 333 & 342 & 348 & 341\\
\midrule
\textsc{Compl.} & SP-SMM & 337 & 441 & 527 & 567 & 456\\
 & \textsc{LipschitzMono} & 688 & 746 & 777 & 861 & 839\\
 & \textsc{SMNN} & 251 & 383 & 431 & 456 & 384\\
 & \textsc{MonoDense} & 164 & 301 & 318 & 317 & 300\\
 & \textsc{MonoNet} & 162 & 299 & 339 & 348 & 318\\
 & \textsc{PMLinear} & 132 & 227 & 238 & 240 & 226\\
 & V-MLP & 117 & 252 & 262 & 265 & 253\\
 & U-MLP & 113 & 216 & 217 & 224 & 215\\
\bottomrule
\end{tabular}
}
\end{table}

\begin{table}
  \centering
  \caption{Class-level service after imitation on the partner benchmark.}
  \label{tab:imitate-fair}
  \small
  \setlength{\tabcolsep}{2pt}
  \resizebox{\columnwidth}{!}{\begin{tabular}{@{}lcccccc@{}}
\toprule
& \multicolumn{3}{c}{Union BC} & \multicolumn{3}{c}{DAggerW}\\
\cmidrule(lr){2-4}\cmidrule(l){5-7}
method & Jain & min.\ rate & served & Jain & min.\ rate & served\\
\midrule
SP-SMM & $0.999\pm0.000$ & $0.424\pm0.001$ & $0.439$ & $1.000\pm0.000$ & $0.443\pm0.001$ & $0.449$\\
\textsc{LipschitzMono} & $0.999\pm0.000$ & $0.423\pm0.003$ & $0.439$ & $1.000\pm0.000$ & $0.442\pm0.002$ & $0.450$\\
\textsc{SMNN} & $0.995\pm0.002$ & $0.393\pm0.010$ & $0.422$ & $0.997\pm0.002$ & $0.412\pm0.005$ & $0.439$\\
\textsc{MonoDense} & $0.997\pm0.001$ & $0.401\pm0.010$ & $0.429$ & $0.995\pm0.003$ & $0.405\pm0.009$ & $0.446$\\
\textsc{MonoNet} & $0.998\pm0.001$ & $0.405\pm0.007$ & $0.427$ & $0.994\pm0.001$ & $0.398\pm0.004$ & $0.447$\\
\textsc{PMLinear} & $0.999\pm0.000$ & $0.420\pm0.003$ & $0.433$ & $0.994\pm0.001$ & $0.397\pm0.002$ & $0.447$\\
U-MLP & $0.974\pm0.000$ & $0.372\pm0.000$ & $0.456$ & $0.979\pm0.005$ & $0.384\pm0.012$ & $0.455$\\
\bottomrule
\end{tabular}
}
\end{table}

\section{Unconstrained and value-only controls}
\label{app:scratch}

Separately, U-MLP and V-MLP are the controls of \cref{sec:benchmarks}. From random initialisation, U-MLP matches FCFS welfare on Partner, Small, Large, and Complementary (\(0.891\), \(0.417\), \(0.559\), \(0.742\)) and is below FCFS on Continuous (\(0.616\pm0.154\) versus \(0.770\)). V-MLP is below FCFS on Partner (\(0.781\pm0.027\)) and above it on the remaining streams.

\paragraph{Misreport search.}
 We freeze each trained score and look for a profitable misreport. We follow the grid used for online electric-vehicle mechanisms~\cite{robu2013jair} and neural auctions~\cite{duetting2019auctions}. We take reported value from the \(21\)-point grid \(\{0,v_{\max}/20,\ldots,v_{\max}\}\) together with the true value, jointly with duration or a single demand coordinate. Utility is that of \cref{sec:incentives}. We write \(R_i\) for the utility gain of agent \(i\), $R_i = U_i(\hat\theta_i,\theta_i,\hat\Theta_{<i}) - U_i(\theta_i,\theta_i,\hat\Theta_{<i}).$
We take \emph{by lie} to be the fraction of audited states at which this gain is positive. A zero in the search means the grid found no improving deviation, not that none exists outside the grid.

After training, the grid returns \(R_i=0\) for every score in \cref{tab:audit-all}, including the controls, so we omit a by-lie column. U-MLP's zero regret coincides with accepting every feasible request and charging \(0\). By contrast, the same search returns \(R_i=0\) for SP-SMM and \textsc{LinearMono} while they remain selective. That is required by \cref{thm:sp} in \cref{sec:invert}; it is a consistency check of the implementation, not an empirical proof.

During training, only the two controls ever have \(R_i>0\). Consequently \cref{tab:audit-worst-all,tab:audit-curve-all} list those controls only; the Partner by-lie rates in \cref{sec:main-result} are the worst-checkpoint column of \cref{tab:audit-worst-all}. Every monotone score stays at zero regret on every stream, including \textsc{SMNN}'s most selective Partner checkpoint (\(18\%\) accepted).

\begin{table}
  \centering
  \caption{After-training misreport search on the partner stream: accepted share, revenue, and welfare.}
  \label{tab:audit-all}
  \small
  \begin{tabular}{@{}lccc@{}}
\toprule
score & accepted & rev. & welf.\\
\midrule
U-MLP & $46\%$ & $0$ & $456$\\
V-MLP & $26\%$ & $223$ & $403$\\
\textsc{LinearMono} & $44\%$ & $168$ & $478$\\
SP-SMM & $45\%$ & $465$ & $491$\\
\textsc{LipschitzMono} & $44\%$ & $475$ & $495$\\
\textsc{SMNN} & $43\%$ & $144$ & $484$\\
\textsc{MonoDense} & $42\%$ & $364$ & $485$\\
\textsc{MonoNet} & $42\%$ & $366$ & $482$\\
\textsc{PMLinear} & $40\%$ & $458$ & $483$\\
\bottomrule
\end{tabular}

\end{table}

\begin{table}
  \centering
  \caption{Worst checkpoint during training on the partner stream, controls only.}
  \label{tab:audit-worst-all}
  \small
  \begin{tabular}{@{}lccc@{}}
\toprule
score & accepted & by lie & max mean regret $/v_{\max}$\\
\midrule
U-MLP & $36\%$ & $43\%$ & $0.351$\\
V-MLP & $44\%$ & $29\%$ & $0.215$\\
\bottomrule
\end{tabular}

\end{table}

\begin{table}
  \centering
  \caption{Maximum mean regret during training as a fraction of \(v_{\max}\), controls only.}
  \label{tab:audit-curve-all}
  \small
  \begin{tabular}{@{}lccccc@{}}
\toprule
score & \textsc{Partner} & \textsc{Small} & \textsc{Large} & \textsc{Cont.} & \textsc{Compl.}\\
\midrule
U-MLP & $0.351$ & $0.076$ & $0.200$ & $0.090$ & $0.450$\\
V-MLP & $0.215$ & $0.050$ & $0.193$ & $0.038$ & $0.342$\\
\bottomrule
\end{tabular}

\end{table}

\section{Service across classes}
\label{app:fair}

Finally, the audit figure in \cref{sec:main-result} plots worst-class service and Jain's index on the partner stream. This section shows the formula and the remaining methods. We write \(\eta_k\) for the fraction of class-\(k\) arrivals that complete, \(k\in\{A,B,C,D\}\), and \(\boldsymbol{\eta}=(\eta_A,\eta_B,\eta_C,\eta_D)\). Jain's index~\cite{jain1984fairness} is
\begin{equation}
  J(\boldsymbol{\eta})=\frac{\bigl(\sum_{k} \eta_k\bigr)^2}{4\sum_{k} \eta_k^2}.
\end{equation}
The index equals \(1/4\) when a single class receives all of the service. We also report the worst-class rate \(\min_k\eta_k\), because \(J\) can remain high while one class is poorly served. Served is the fraction of all arrivals that complete, so a rule that allocates nothing cannot top the table: \(J\) is undefined on the zero vector. FCFS is a reference for class-level differences caused by the arrival process, because it makes no value-based selection.

\Cref{tab:fair-all} lists \(J\), the worst-class rate, and served fraction after RL from random initialisation. U-MLP's post-training rule accepts every feasible request (\cref{app:scratch}), so its class mix matches FCFS and we omit its row. The worst-served class is always a filler. V-MLP starves class C in four of five seeds; those four seeds have \(J=0.50\), while its overall mean is \(0.595\pm0.095\) because the fifth seed has FCFS-like service. None of the networks monotone in every reported quantity starves a class in these runs.

\begin{table}
  \centering
  \caption{Per-class service after training from random initialisation. U-MLP matches FCFS and is omitted.}
  \label{tab:fair-all}
  \small
  \begin{tabular}{@{}lccc@{}}
\toprule
method & Jain $J(\boldsymbol{\eta})$ & min.\ rate & served\\
\midrule
FCFS & $0.974$ & $0.372$ & $0.456$\\
SP-SMM & $1.000\pm0.000$ & $0.440\pm0.002$ & $0.446$\\
\textsc{LipschitzMono} & $0.997\pm0.001$ & $0.413\pm0.013$ & $0.436$\\
\textsc{LinearMono} & $0.998\pm0.001$ & $0.414\pm0.003$ & $0.436$\\
\textsc{SMNN} & $0.997\pm0.001$ & $0.403\pm0.007$ & $0.426$\\
\textsc{MonoDense} & $0.995\pm0.003$ & $0.384\pm0.016$ & $0.422$\\
\textsc{MonoNet} & $0.996\pm0.001$ & $0.386\pm0.006$ & $0.417$\\
\textsc{PMLinear} & $0.985\pm0.004$ & $0.335\pm0.017$ & $0.405$\\
V-MLP & $0.595\pm0.095$ & $0.074\pm0.074$ & $0.252$\\
\bottomrule
\end{tabular}

\end{table}

%%%%%%%%%%%%%%%%%%%%%%%%%%%%%%%%%%%%%%%%%%%%%%%%%%%%%%%%%%%%%%%%%%%%%%%%

\end{document}